\documentclass[twocolumn, switch]{article} % Method A for two-column formatting

\usepackage{preprint}

\usepackage{amsmath, amsthm, amssymb, amsfonts}

\usepackage[numbers,square]{natbib}
\usepackage[utf8]{inputenc}	% allow utf-8 input
\usepackage[T1]{fontenc}	% use 8-bit T1 fonts
\usepackage{xcolor}		% colors for hyperlinks
\usepackage[colorlinks = true,
            linkcolor = purple,
            urlcolor  = blue,
            citecolor = cyan,
            anchorcolor = black]{hyperref}	% Color links to references, figures, etc.
\usepackage{booktabs} 		% professional-quality tables
\usepackage{nicefrac}		% compact symbols for 1/2, etc.
\usepackage{microtype}		% microtypography
\usepackage{lineno}		% Line numbers
\usepackage{float}			% Allows for figures within multicol
\usepackage{multicol}		% Multiple columns (Method B)
\usepackage{tabularx}
\usepackage{adjustbox}
\usepackage{rotating}
\usepackage{lipsum}		%  Filler text
\usepackage{mathabx}
\usepackage{mdwmath}
\usepackage{mdwtab}
\usepackage{array}
\usepackage{eqparbox}
\usepackage{newfloat}
\DeclareFloatingEnvironment[name={Supplementary Figure}]{suppfigure}
\usepackage{sidecap}
\sidecaptionvpos{figure}{c}

\usepackage{titlesec}
\titlespacing\section{0pt}{12pt plus 3pt minus 3pt}{1pt plus 1pt minus 1pt}
\titlespacing\subsection{0pt}{10pt plus 3pt minus 3pt}{1pt plus 1pt minus 1pt}
\titlespacing\subsubsection{0pt}{8pt plus 3pt minus 3pt}{1pt plus 1pt minus 1pt}

\title{Design and Evaluation of A Cyber-Physical Resilient Power System Testbed}

\usepackage{xwatermark}
\newwatermark[firstpage,color=gray!60,angle=90,scale=0.32, xpos=-4.05in,ypos=0]{\href{https://doi.org/}{\color{gray}{Publication doi}}}
\newwatermark[firstpage,color=gray!60,angle=90,scale=0.32, xpos=3.9in,ypos=0]{\href{https://doi.org/}{\color{gray}{Preprint doi}}}
\newwatermark[firstpage,color=gray!90,angle=0,scale=0.28, xpos=0in,ypos=-5in]{*correspondence: \texttt{email@institution.edu}}

\usepackage{authblk}

\author[1]{Abhijeet Sahu}
\author[2]{Patrick Wlazlo}
\author[1]{Zeyu Mao}
\author[1]{Hao Huang}
\author[2]{Ana Goulart}
\author[1]{Katherine Davis}
\author[3]{Saman Zonouz}
\affil[1]{Electrical and Computer Engineering, Texas A\&M University, College Station, US}
\affil[2]{
Electronic Systems Engineering Technology, Texas A\&M University, College Station, US}
\affil[3]{Electrical and Computer Engineering, Rutgers University, New Jersey, US}

\begin{document}

\twocolumn[ % Method A for two-column formatting
  \begin{@twocolumnfalse} % Method A for two-column formatting
  
\maketitle

\begin{abstract}
A power system is a complex cyber-physical system whose security is critical to its function. A major challenge is to model and analyze its communication pathways with respect to cyber threats. To achieve this, the design and evaluation of a cyber-physical power system (CPPS) testbed called Resilient Energy Systems Lab (RESLab) is presented that captures realistic cyber, physical, and protection system features. RESLab is architected to be a fundamental tool for studying and improving the resilience of complex CPPS to cyber threats. The cyber network is emulated using Common Open Research Emulator (CORE) that acts as a gateway for the physical and protection devices to communicate. The physical grid is simulated in the dynamic time frame using PowerWorld Dynamic Studio (PWDS). The protection components are modeled with both PWDS and physical devices including the SEL Real-Time Automation Controller (RTAC). Distributed Network Protocol 3 (DNP3) is used to monitor and control the grid. Then, exemplifying the design and validation of these tools, this paper presents four case studies on cyber-attack and defense using RESLab, where we demonstrate false data and command injection using Man-in-the-Middle and Denial of Service attacks and validate them on a large-scale synthetic electric grid.
\end{abstract}
%\keywords{First keyword \and Second keyword \and More} % (optional)
\vspace{0.35cm}

  \end{@twocolumnfalse} % Method A for two-column formatting
] % Method A for two-column formatting

%\begin{multicols}{2} % Method B for two-column formatting (doesn't play well with line numbers), comment out if using method A

%%%%%%%%%%%%%%%  Main text   %%%%%%%%%%%%%%%
% \linenumbers

\section{Introduction}\label{sec:intro}
The electric grid is transitioning to a smarter grid that employs advanced communication technologies. With advanced computing and communications, cyber-security has proven to be a critical issue in power transmission, generation, and distribution systems. Cyber adversaries can modify or create
data that can impact the grid's normal operation and potentially destabilize its operating point causing cascading failures.
Earlier this year, an unidentified threat successfully compromised the 
administrative systems of the European Network of Transmission System Operators for Electricity (ENTSO-E), with the potential to compromise 42 transmission system operators (TSOs) across 35 member states in Europe \cite{entso}. Other attacks are also widely known such as the Ukraine attacks \cite{ukraine_2015}, where an attacker targeted three distribution units to cause a power outage after intruding into the Supervisory Control and Data Acquisition (SCADA) system. Attacks like Pivnichna \cite{pivnichna} caused a power outage, while Stuxnet \cite{stuxnet} allowed control of programmable logic controllers (PLCs), by overspeeding the centrifuges in a nuclear plant.

It is necessary to propose defense mechanisms for such zero-day attacks. The use of firewalls, intrusion detection systems, and intrusion prevention systems is important, but these tools may not work efficiently on stealthy coordinated attacks. Hence, we need to employ the latest tools and techniques to make solutions that are more intelligent and capable of detecting complex attacks. Machine learning, including deep learning, or even artificial intelligence, offer advantages that can aid cyber and physical attack detection and localization. These techniques are data-intensive, where more data typically provides a better solution. One way to generate those real-time data sets is to mimic those attacks and defense mechanisms using a testbed.  

This paper presents our Resilient Energy Systems Laboratory (RESLab) testbed that forms an environment for researchers and stakeholders to understand the impact of cyber-attacks and validate their defenses. It provides a platform to evaluate how the power and communication networks perform together based on real-world systems and events, including communication protocols, operations, and latency requirements. It allows other researchers to develop and test intrusion detection tools for defending and mitigating real cyber attacks.  

These are the major contributions of this paper:
\begin{enumerate}
\item To introduce RESLab, a cyber-physical power system testbed that is designed to study resilience problems and solutions in large-scale power systems RESLab is a mix of emulators, simulators, and real devices that allow us to evaluate multi-stage cyber threats to the power system.

\item To model realistic data flows in the RESLab testbed using a large-scale exemplar power system based on utility architecture. This enables us to implement and validate scenarios impacting grid resilience such as false data and command injection uses cases.

\item To compare RESLab with other testbeds and present how RESLab is able to implement and validate realistic use cases for grid cyber-resilience.  

\item To implement Denial of Service (DoS) and Man-in-the-Middle (MiTM) attacks and validate them by studying their impacts on the normal power system operation. 

\item To provide a platform for data collection and visualization by integrating monitoring tools such as Packetbeat and Zabbix, Snort for intrusion detection, and a cyber-physical energy management system application.

\end{enumerate}

This paper is organized as follows. In Section~\ref{related_works}, we evaluate testbeds that model a cyber-physical power system and allow real-time experiments. Section~\ref{cypres_architecture} presents the architecture and components of RESLab. The threat model for this work is presented in Section~\ref{cyber_threats}. Section~\ref{use_cases} demonstrates its implementation in RESLab with four use cases. Then, we present the analysis of the attacks and their impacts on the physical system. The results are analyzed in Section~\ref{results}.

\section{Cyber-Physical Power System Testbeds}
\label{related_works}
In this section, we first review testbeds that focus on investigating vulnerability of power critical infrastructure, including their challenges and limitations.  
%Relevant prior 
Previous works range from applications in wide-area protection and monitoring in transmission and generation, to distributed energy resources (DERs), to microgrids and distribution systems, and to operation domains such as Energy Management Systems (EMS) and Distribution Management Systems (DMS).  The comparisons in this section consider ease of deployment and troubleshooting, design complexity, and cost of implementation. Then, we motivate RESLab, outlining its contributions, unique features, and how it fills existing gaps.  %Further, in the next section, we detail each component of RESLab through its architecture. 

\subsection{Testbeds and platforms}
%The creation of a cyber-physical testbed typically involves designing an architecture comprised of cyber and physical components. 
A cyber-physical testbed architecture is implemented by networking together simulators, emulators, and hardware.  The quality of a cyber-physical testbed is measured by its success in advancing the research and applications that it supports.  
%Characteristics present in high-quality cyber-physical testbeds 
It includes a platform enabling communication between components, a system for data collection, aggregation, visualization, and a way of executing and evaluating cyber security incidents against the system under study.

We present all the reviewed testbeds in Table~\ref{tab:reveiw_testbed} based on the power and cyber simulators, communication protocols, software and devices, system level, intrusions type, and the application of DERs. 

\subsection{Network representation}
Various testbeds such as ~\cite{pan2015developing, adhikari2016wams,poudel2017real,yang_2017,igor_2010, idaho_testbed} focus on evaluating impact for physical use cases (e.g., cyber-attacks on power flows, loss of load or synchronism, protection systems, 
%DERs, 
and microgrids) while using networking hardware.   Other testbeds ~\cite{bo_karen, austin_2016,malaz_2011, hossein_2015, abhijeet_testbed} use network simulators, but the network simulator's primarily focus is on %objectives such as 
communication algorithms such as 
%TCP 
congestion control or bandwidth allocation schemes.
%, and queueing theories.  
For performing cyber-physical studies in a testbed, network emulation is the preferred alternative to simulation.  While a simulator 
%is designed to perform tasks to 
demonstrates the behavior of a network, an emulator functionally replicates the behavior of the network. Though some simulators provide features such as System-in-The Loop (SITL)~\cite{bo_karen,sitl2011} and Tap bridge ~\cite{abhijeet_testbed} to integrate external devices or virtual machines (VMs), those features are not scalable to large %cyber-physical 
systems.  

Virtualization and emulation enable scalability. In~\cite{vsphere}, the vSphere ESXi virtualization environment is used to simulate network-based attacks; however, these scenarios are not specific to SCADA and focus on web browsing and file transfers. The vSphere environment is reset 
%the simulation environment 
before each simulation test run; this provides built-in mitigation against any damage resulting from an attack simulations.   

The selection of a specific platform requires design decisions that are based on trade-offs in cost and accuracy. Network devices such as firewalls and routers can be included in the design, 
%Such components support accuracy 
but are expensive for a large scale CPS. Network simulators such as Opnet, Omnet, and Network Simulator-3 (NS-3) can be cost-effective, but they do not provide a platform for real-time data processing 
%that are encapsulated in 
of industrial protocols such as IEC 61850, Distributed Network Protocol version 3 (DNP3), inter-control center communications protocol (ICCP), and Modbus.

The Common Open Research Emulator (CORE) provides a platform to run different applications,
%within network emulation, 
such as iptables for firewall, Snort for intrusion detection, and services such as Secure Shell (SSH) for remote access. CORE is used for emulating smart grid networks in~\cite{core2}, where the authors compare existing works of co-simulation and discover CORE to be suitable for large-scale simulations. 
%They also find that applications in the testbed are easily deployable to embedded devices. %AG: this sentence is not very clear. Which applications? Which embedded devices? Is this about adding real hardware devices?
Similarly, in~\cite{core1} an Army microgrid is simulated, and its communication network is emulated with CORE, where results demonstrate the ability to implement cyber intrusions and evaluate the impact on the microgrid. However, testbeds that use CORE
%for network emulation 
have not yet integrated network monitoring tools. %in their platforms. 

Sandia National Laboratories (SNL) offers a tool for launching and managing virtual machines (VMs) to emulate large scale networks~\cite{raybourn2018zero} called Minimega. It supports Virtual Local Area Networks (VLANs) with configurable bandwidth and quality of service (QoS); thus, it can be used with a router’s operating system and Linux kernels to emulate a communication network. %infrastructure.
%with high fidelity. (AG: high-fidelity shows twice in this paragraph)
Additionally, a power system and cyber co-simulation environment called SCEPTRE~\cite{sceptre} is being developed by SNL to allow high-fidelity simulation of SCADA protocols with hardware-in-the-loop such as
%the integration of 
PLCs and remote terminal units (RTUs), using Minimega to manage the VMs. 

A Hierarchical Engine for Large-scale Infrastructure Co-Simulation (HELICS) is developed by National Renewable Energy Laboratory for large-scale co-simulation, using off-the-self power system, and communication markets~\cite{helics}.
%and end-use tools - AG: what are end-use tools?
This framework integrates discrete-event simulators, such as NS-3,
%and CORE (is CORE discrete-event?)
and time-series simulations such as for power flows. %We have utilized their framework for expanding our testbed to emulate individual outstations within CORE for traffic segregation. The details of its integration will be presented in our future work.

%\textcolor{red}{[HELICS? PNNL testbed?]} 
% Sandia National Laboratories (SNL) currently offers a tool for launching and managing virtual machines (VMs) to emulate large scale networks~\cite{raybourn2018zero} called Minimega. It supports Virtual Local Area Networks (VLANs) with configurable bandwidth and quality of service (QoS); thus, it can be used with a router’s operating system (OS) and Linux kernels to emulate a physical communication network infrastructure with high fidelity. Additionally, a power system and communication network co-simulation environment called SCEPTRE~\cite{sceptre} is being developed by SNL that allows high-fidelity simulation of SCADA protocols, such as Modbus, TCP, DNP3, and IEC 61850, alongside hardware-in-the loop features with integration of PLCs and RTUs. SCEPTRE uses Minimega to manage the VMs.
% \as{I am not sure shall we discuss about Minimega or HELICS in detail here}

% A Hierarchical Engine for Large-scale Infrastructure Co-Simulation (HELICS) is developed by National Renewable Energy Laboratory for 
% %a very 
% large-scale co-simulation, utilizing off-the-self power system, communication, markets, and end-use tools~\cite{helics}. This framework supports integration of discrete event simulations such as NS-3 and CORE as well as time-series simulations such as power flows. We have utilized their framework for expanding our testbed to emulate individual outstations within CORE for traffic segregation. The details of its integration will be presented in our future work.
%on securing the grid with a reconfigurable cyber network. 

\subsection{Power system representation}
Power system simulators such as real-time digital simulator (RTDS), OpalRT, or Typhoon have been used in several testbeds~\cite{hong2015cyber,liu2015analyzing,kezunovic2017use,bo_karen,poudel2017real,adhikari2016wams,ashok2015experimental,papaspiliotopoulos2015hardware,yang_2017,idaho_testbed,iowaTestbed}. These expensive hardware solutions are essential for experiments
%focused 
on electromagnetic transients or 
%for research in 
power electronics, DERS, or microgrids.  
PowerWorld Simulator and Dynamic Studio (PWDS) provide solutions for large-scale power system modeling in the steady state and transient stability time frames ~\cite{DS,overbye2017interactive,w4ips}. 
The testbeds that use hardware devices such as SEL relays, phasor measurement units (PMUs), or RTACs for their experiments face challenges in understanding vendor-specific
%needing to obtain or develop solutions and tools that understand 
industrial control system (ICS) protocols. The testbed solutions in ~\cite{hong2015cyber,liu2015analyzing,kezunovic2017use,johnson2018interconnection,igor_2010,papaspiliotopoulos2015hardware} use physical hardware and consider %vendor-specific 
ICS protocols such as IEC 61850, C37.118 in their experiments.

\subsection{Operational technology protocols}
Widely-known tools in %traditional 
information technology (IT) security, such as Ettercap and Metasploit frameworks, 
%also face difficulty because they %tend to not be 
are not
tailored to operational technology (OT) or attacks to
%specifically geared toward 
power systems. For example, a MiTM attack would need to be specific to SCADA protocols, and representing such attacks is imperative for defense. Incorporating attacks and defense into a testbed requires knowledge of the protocols as it involves inspecting and modifying packets.
%sniffing and modifications. 
Vendor-specific protocols that are not open source are challenging to incorporate and evaluate in such environments. 
%Testing solutions for cyber-attack resilience of power systems requires a safe proving-ground.  

\subsection{Constraints on use cases}
Existing testbeds often lack strong demonstrations of cyber intrusions. Several existing testbeds ~\cite{hong2015cyber,johnson2018interconnection,thornton2017internet,piesciorovsky2017fuse,tamara_pnnl,austin_2016,behrouz_2019,stifter2018real} mention implementation of cyber intrusions in their platforms but do not clearly demonstrate specific use cases including how the intrusions are performed.  Some testbeds that use MiTM attacks ~\cite{kezunovic2017use,poudel2017real,ashok2015experimental,adhikari2016wams,yang_2017} do not show how those attacks are incorporated. For example, a MiTM attack can be performed in different ways such as address resolution protocol (ARP) cache poisoning, internet protocol (IP) spoofing,
%or SQL injections 
or hypertext transfer protocol (HTTP) session hijacking. Some testbeds such as ~\cite{bo_karen,yang_2012} use Ettercap or Metasploit frameworks but have limitations in carrying out goal-oriented MiTM cyber-attacks. 

 Realistic use case support is a major feature of 
 %an ideal 
 a testbed. For example, extensive research has been proposed on defense against FDI attacks, where state-of-the-art methods adopt linear algebra and deep learning ~\cite{se_attack,ozay2016machine}. However, works that address FDI attacks tend to make unrealistic assumptions on the adversary's knowledge and capabilities. 

\subsection{Motivation for RESLab}
RESLab aims to facilitate academic research and to bridge theory
%theoretical results 
to practice through collaboration with
industry and academia. 
%stakeholders.  
Therefore, it must exemplify these qualities: 
%Design is able
(1) Ability to 
%demonstrate success in the 
validate domain-specific use cases; (2) Ability to mimic and reflect the real-world complexity of industrial systems by incorporating hardware devices; 
%achieving this involves hardware features; 
%(3) Ease of deployment
%of prototypes to in industry; (this is mentioned when we say transfer the testbed to industry
(3) %Enabling other research %fraternity
%collaborations and to 
Support high-impact research activities; (4) Cost effectiveness; 
%(6) Ability to support high-impact research activities; 
(5) Design for fast verification of results; (6) Ability to transfer results to power system industry; (7) Ability to transfer the testbed itself to 
%power system 
industry or other researchers; and (8) Ability to serve as an educational platform.
%including 
%by incorporating multi-user experimental facilities. 

Because RESLab is a cyber-physical testbed, these features are also important: (1) It is imperative that the testbed allows the development and evaluation of cross domain (e.g., cyber, physical, protection system) analyses including identification of cyber threats that impact power systems; (2)	Ability to implement and validate realistic threats using real-time simulated models and data as well as offline models; (3) Ability to evaluate cyber-physical contingency analysis involving simulating and patching vulnerabilities and their impact on risk; (4) Ability to 
%network
connect
simulators, emulators, and physical components.

These RESLab features fulfill the gaps in existing testbeds:
\begin{enumerate}
    \item \textit{Virtualization and emulation:} RESLab uses vSphere and CORE for virtualization,
    %purpose at different levels 
    including power system simulation in a dedicated VM, 
    and to operate network components such as an IDS and firewalls.
    %within a Berkeley Software Distribution (BSD) jails inside CORE. 
    \item \textit{Open source protocols and industry integration:} RESLab’s implementation of OpenDNP3 follows the IEEE 1815 standard~\cite{open_dnp3},
    %and is not vendor-specific, 
    hence convenient for other researchers to replicate the experiments. 
    RESLab uses PWDS~\cite{DS} to emulate the DNP3 outstations in real-time, making the solution easily deployable to industry. 
    \item \textit{Use case realism support:} 
    %RESLab extends the state-of-the-art with respect to the use case shortcomings described above because
    RESLab addresses use case shortcomings by %specifically designing 
    supporting multi-phase cyber intrusions as presented by the joint report of NERC, E-ISAC, and SANS-ICS~\cite{cyber_killchain}. Other solutions fail to address early attack stages.
    %and tend to ignore these stepping stone events that would be in place at a real utility.
%    Specifically, as shown in Section~\ref{use_cases}, RESLab is used to implement a cache poisoning and MiTM script that is customized to achieve specific use cases such as false data injection (FDI) against state estimation and false command injection (FCI) to cause contingencies and cascading failures. 
Our testbed enables a complete representation of FDI and  FCI attack vectors in a realistic environment from early-stage environment. %RESLab enables the implementation of early-stage mechanisms; for example, we can emulate an intruder's initial action to sniff traffic to predict system topology before injecting FDI attacks. 
 %By providing a realistic utility environment and data sources, RESLab supports developing defense mechanisms around existing power system functions including state estimation in energy management systems.
    
    \item \textit{Large scale system cyber-physical analysis:} RESLab enables research to develop entirely new systems. For example, it is supporting the use case of a cyber-physical EMS. Such a new system needs to support algorithms that %consider cyber states to implement and
    enable cyber-physical state estimation. %formulations. 
    Its integration and application of large-scale realistic cyber-physical models, including a synthetic test case on the Texas footprint that includes power~\cite{synthetic} and communication~\cite{synthetic_comm} systems, with balancing authorities and market participants strengthens the test case to mimic a realistic cyber-physical power system. 
    %with detailed data pipeline and architecture.

% Additionally, the major benefit of RESLab goes beyond securing existing platforms: RESLab enables research to develop entirely new systems. For example, it is supporting the use case of a cyber-physical energy management system (EMS). Such a new system needs to support algorithms that consider cyber states to implement and validate cyber-physical state estimation formulations.  In support of a cyber-physical EMS prototype, RESLab also features graphical user interfaces (GUIs) that the team developed for control and monitoring using the open DNP3 python and pyQT package. 

% One of the key features of RESLab is its integration and application of large-scale realistic cyber-physical models, including a synthetic test case on the Texas footprint that includes power~\cite{synthetic} and communication~\cite{synthetic_comm} systems, with integration of balancing authorities, market
% participants such as Load Serving Entity (LSE), Transmission and Distribution Serving Entity (TDSE), etc. This test case is designed to enable experimentation though mimicking a realistic cyber-physical power system data pipeline and architecture.  This architecture models the types of communications, protocols, firewall policies, etc., that are involved in power system operation. 
    
    \item \textit{Transferability and interoperability:} The virtualization in RESLab makes the migration to other %virtualization 
    platforms such as VirtualBox and VMware simple. This approach makes the testbed cost-effective compared to testbeds that use RTDS or OpalRT.
    %of our platform is a crucial feature of RESLab.  The approach in RESLab of creating a cyber-physical testbed using virtualization is cost-effective compared to using RTDS, OpalRT, or physical networking devices which are infeasible to mimic large-scale power networks. The virtualization environment also enables RESLab to be easily deployable, i.e., our environment can be migrated to other virtualization platforms such as VirtualBox and VMware Fusion. The VMs are deployable to other researchers to mimic the same experiments that we perform. Our testbed is also accessible to individual and industry stakeholders who may be unable to invest significantly in hardware.
    \item \textit{Dataset management:} Few datasets for cyber physical testbeds are publicly available. RESLab provides such a platform that aggregates real-time traffic and power data along with IDS alerts and
    %to provide 
    %such 
    %solutions using the Elasticsearch database. 
    %It 
    enables integration of third-party tools including visualization and data analytics.
    %, a cloud-based NoSQL storage solution that is widely used by industry. By adopting this approach to storage, integration of third-party applications (e.g., for security information and event management (SIEMS) and visualization) becomes convenient.
\end{enumerate}

%In Section~\ref{cypres_architecture}, we discuss each component of RESLab in detail.

% The table for comparison of the test-beds
\begin{sidewaystable*}[!htbp]

%\begin{adjustbox}{scale=0.95,center}
    \centering
\caption{Review of Cyber-Physical Power System Testbeds}

    \label{tab:reveiw_testbed}
\begin{adjustbox}{max width=0.65\textwidth, center} %{scale=0.7}
%\begin{adjustwidth}{-20cm}{}
\hskip-3cm
\begin{tabularx}%[b]
%\hspace{-1cm}
%{ C{1.1cm} C{1.1cm} C{1.1cm} C{1.1cm} C{1.1cm} C{1.1cm}C{1.1cm} C{1.1cm} C{1.1cm}}    
%{\textwidth}{| m{15em}|*{7}{>{\RaggedRight\arraybackslash}X|}}
%{\textwidth}{| m{15em}|*{7}{>{\RaggedRight\arraybackslash}l|}}
%{\textwidth}{| m{14em}|m{4em}|m{5em}|m{4em}|m{5em}|m{4em}|m{5em}|m{4em}|}
%{\columnwidth}{|m{1.5em}|m{10em}|m{10em}|m{14em}|m{14em}|m{10em}|m{12em}|m{2.5em}|}
{\columnwidth}{|m{3.5em}|m{9em}|m{10em}|m{14em}|m{14em}|m{10em}|m{12em}|m{2.5em}|}
%\begin{tabular}{ C{1.1cm} C{1.1cm} C{1.1cm} C{1.1cm} C{1.1cm} C{1.1cm}C{1.1cm} C{1.1cm} C{1.1cm}}  
\hline
\textbf{Ref} & \textbf{Power Simulators} & \textbf{Cyber Simulators} & \textbf{Communication Protocols} &  \textbf{Devices and Softwares}  &   \textbf{System Level}  &   \textbf{Intrusions} & \textbf{DER} \\
    \hline
    \hline
\cite{hong2015cyber}
 & RTDS & No &IEC 61850 &  Relays, IEDs, Gateway  &   Substation  &   No & No \\
    \hline
\cite{liu2015analyzing}
 & RTDS & Network Simulator-3/DeterLab & PMU/C37.118
 &  PMUs, phasor data concentrator (PDC), GPS clock &   Transmission &   DoS, MiTM & No \\
    \hline
\cite{kezunovic2017use}
 & RTDS, Opal-RT & Wide Area Communication Emulator&C37.118 &  PMUs, SDN, RTAC, PDC, Industry-grade SCADA  &   Transmission  &   MiTM & No \\
    \hline
\cite{poudel2017real}
 & OPAL-RT & Real Network/SDN& DNP3 & SEL 351 &   Transmission  &   DoS and coordinated physical control & No \\
    \hline
\cite{pan2015developing, adhikari2016wams}
 & RTDS & Real Network/SDN&MODBUS/TCP, IEEE C37.118 & PMU, PDC, relay, Industry software from SEL, GE, Snort, Wireshark &   Transmission  &   Aurora, DoS & No \\
    \hline
\cite{ashok2015experimental}
 & RTDS & & DNP3 & Scapy &   Transmission  &   MiTM & No \\
    \hline
\cite{ johnson2018interconnection}
 & Typhoon HIL 602 & No & IEC 61850 & Digitl Signal Processor (DSP), FPGA,  SunSpec System Validation Platform, Inverter and converter &   Distribution  &   No & Yes \\
    \hline
\cite{thornton2017internet}
 & Powersim & Dynamic Link Library (DLL) & Wifi with MODBUS register, SSH & IoT, relay &   Distribution  &   No & Yes \\
    \hline
\cite{piesciorovsky2017fuse}
 & PowerWorld, MATLAB, RT-LAB, OP5600 & &  & Relay &   Distribution  &   No & Yes \\
    \hline
\cite{papaspiliotopoulos2015hardware}
 & RTDS &No &  &PLC, Relays &   Distribution  &   No & Yes \\
    \hline
\cite{stifter2018real}
 & Opal-RT and FPGA & No &C37.118, Precision time protocol (PTP)  & &   Distribution  &   No & No\\
    \hline
\cite{bo_karen}
 & RTDS & Opnet & MODBUS  & LibModbus, OPNET, RTLAB &   Transmission (Attack on Static Var compensator controller)  &   MiTM & No\\
    \hline
\cite{yang_2017}
 & RTDS & real network & IEC 61850  & ITACA IDS tool &   Transmission (Attack on Static Var compensator controller)  &   32 types of attacks, MiTM & No\\
    \hline
\cite{tamara_pnnl}
 & Power World DS & Wide area communication Emulation & DNP3 and GOOSE  & NI CRIO, SEL 421, SEL 651R, SEL 734B &   Transmission  &   No & No\\
    \hline
\cite{idaho_testbed}
 & RTDS & SDN based switch, firewall & IEC C37.118, IEC 61850, DNP3  & Self developed SCANVILLE, RADICL &   No specific use case  &  RADICL for cyber attacks& No\\
    \hline
\cite{igor_2010}
 & Real hardware & Real hardware& &  & No specific use case  &  RADIUS DoS, ICT Worm infection, MALWARE infection, Phishing attack, DNS poisoning& No\\
    \hline
\cite{malaz_2011}
 & PowerWorld, MATLAB, RT-LAB, OP5600 &	OPNET &MODBUS- RSIm & &   No specific use case  &  DoS (compromised HMI), SYN ACK flooding	& No\\
    \hline
\cite{nicol_overbye}
 & PowerWorld &	RINSE (network emulator) &MODBUS-TCP & &   Transmission  &  DDoS attack	& No\\
    \hline
\cite{austin_2016}
 & OPAL-RT & OMNET++ & MODBUS& Ametek MX-45 bi-directional grid simulator for amplification&   Microgrid (IEEE 13)  &  No	& No\\
    \hline
\cite{hossein_2015}
 & OPAL-RT & No simulator or emulator & DNP3 and MODBUS, C37.118 & 	Labview CRIO, OSIsoft's PI-Server &   No  &  No	& No\\
    \hline
\cite{behrouz_2019}
 &OPAL RT and Typhoon HIL& & No & 	Xilinx Virtex 6 FPGAs &   microgrid: specific to generator control  &  No & No\\
    \hline
\textbf{RESLab}
 &\textbf{PowerWorld DS} & \textbf{CORE} & \textbf{DNP3} & \textbf{RTAC, Snort, Packetbeat, OpenDNP3} &   \textbf{Transmission}  &  \textbf{MiTM and DoS} & \textbf{No}\\
    \hline
    
\end{tabularx}
%\end{adjustwidth} 
\end{adjustbox}

%\end{tabular}
%\vspace{-2cm}
\end{sidewaystable*}

%%%%%%%%%%%% Section 3%%%%%%%%%%%%%%%%%%%%%%%%%%%%%%%%%%%%%%%%%

\section{RESLab Cyber-Physical Testbed Architecture}
\label{cypres_architecture}

\begin{figure}[h]
\centering
\includegraphics[scale=.55]{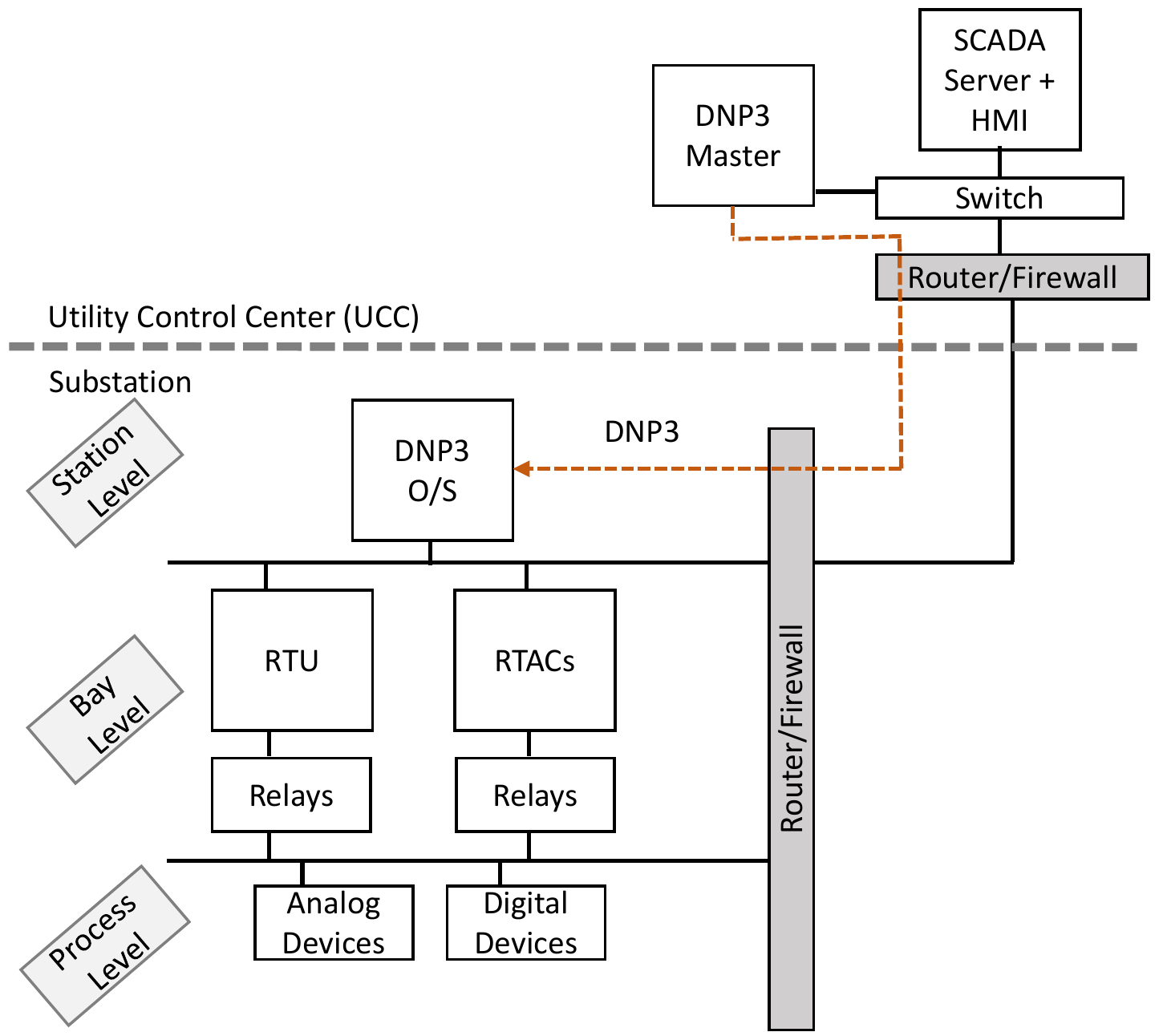}
\caption{Power system cyber-physical architecture with one substation and a utility control center. 
}
\label{fig:high_level_architecture}
\end{figure}

RESLab is designed to reflect realistic power and cyber components based on the synthetic electric grid model on the Texas footprint~\cite{synthetic}, where its communication model
%, or cyber topology, 
is introduced in~\cite{synthetic_comm}.
%(Section~\ref{usecaserealism})
Fig.~\ref{fig:high_level_architecture} presents a high-level view of the RESLab architecture, showing an example of one substation and one utility control center (UCC), with their power system cyber-physical components and data flows. More detailed data flows which include balancing authorities and demilitarized zones (DMZs) are presented in our prior work on 
%creating 
firewall policies that follow NERC standards~\cite{firewall_paper}. 

In the simplified model of Fig.~\ref{fig:high_level_architecture}, the main data flow depicted is DNP3 traffic, which is initiated at the UCC, where there is a DNP3 master and a SCADA server that act as our central control and human machine interface (HMI) applications.  At the substation level, there is a DNP3 outstation (DNP3 O/S) which has the data from the field devices that the UCC needs to monitor.

The RTUs, RTACs, and relays monitor the system status, collect data, and control physical devices such as circuit breakers, which we call digital devices as they have only two states, or generators and load which we call analog devices. The RTUs and RTACs can control generators’ output and affect loads. The relays can trip a circuit breaker to isolate a faulted circuit. The data from the \textit{Process Level} is concentrated in RTUs and RTACs and then transferred to the \textit{Substation Level}. Within the UCC, the DNP3 master collects the information from each substation for a complete view to understand and control the system. In \cite{huang2018extracting}, a typical structure of data concentration and engineer access is presented using the SEL RTAC in SCADA systems, where the RTACs in substations communicate with RTACs in UCC and EMS for data collection and control.
%with communication protocols. 

\begin{figure}[!htb]
  \centering

\includegraphics[width=1\linewidth]{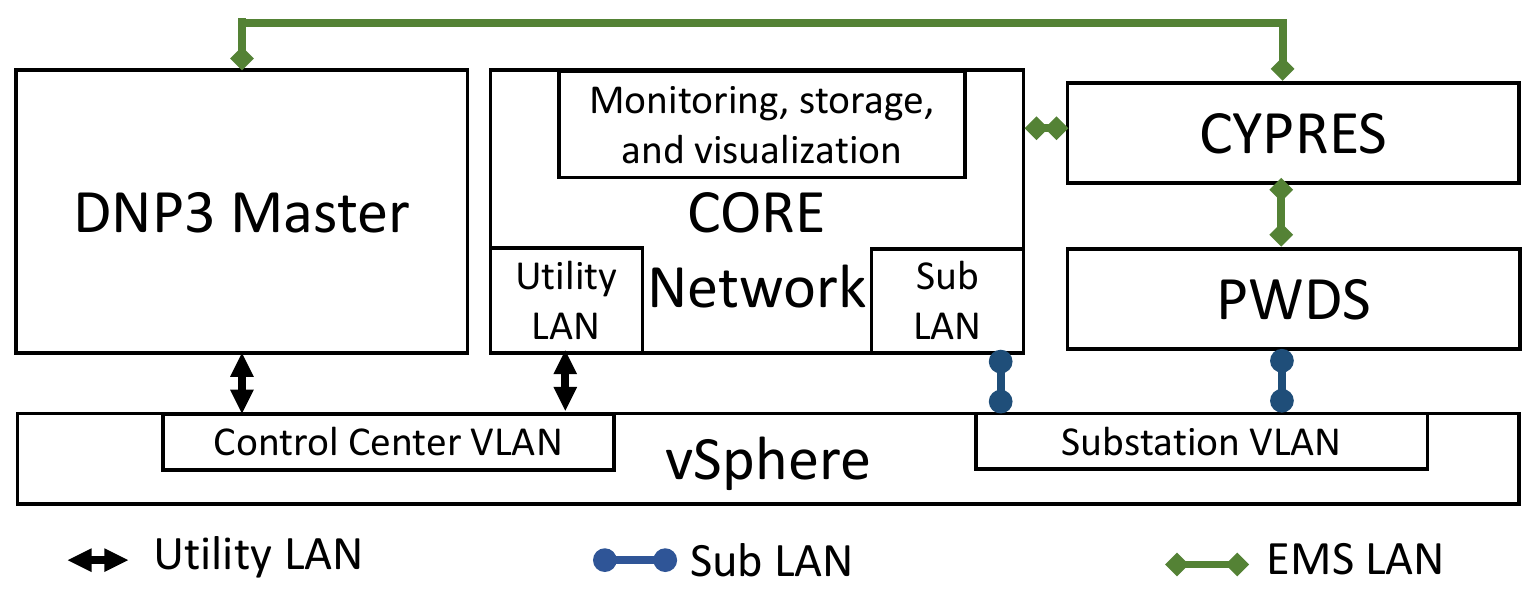}
  \caption{The logical connections between the VMs hosted in RESLab. 
  }
  
  \label{fig:logical_CORE_network}
\end{figure}

The UCC and substations are in different  locations, and in RESLab they are interconnected by an IP network, but they can also be connected through a serial link.
At each location we have one router: the substation router and the UCC router, which also act as firewalls because they are configured to allow only DNP3 packets and block unwanted traffic. 
Fig.~\ref{fig:logical_CORE_network} illustrates how RESLab follows this data flow pipeline and incorporates real-time power system simulation using PWDS, a physical SEL RTAC, an OpenDNP3 master application, and an emulated communication network using CORE.  

In the testbed, PWDS acts as a collection of DNP3 outstations connected to the substation’s control network (shown as Sub LAN in Fig.~\ref{fig:logical_CORE_network}). %Physical devices such as SEL’s RTAC act as DNP3 master in the left. 
The emulated DNP3 master and SEL RTAC are housed in a control center network to represent software- and hardware-based
control platforms. Each of the emulated components are hosted in a virtual machine management environment, named vSphere. The vSphere environment allows for the creation and management of a large number of VMs. In RESLab, connections between emulated and physical components are made to scale the network depending on the use case. Next, the purpose and functionality of each testbed component is analyzed.

\subsection{Cyber network emulation: CORE} 
 RESLab uses CORE, 
 which is an open source network emulator published by the U.S. Naval Research Laboratory. The software allows the creation of several BSD jails, which are similar to Linux containers, that can be connected to emulate realistic communication networks. 
 These containers are used to emulate routers, firewalls, personal computers, and Linux servers in the communication network. 
 CORE can also tap into the hosts’ Ethernet connections to connect with external networking devices and VMs housed within vSphere.

In our testbed, CORE is hosted as one of the VMs, with each of its virtual network interfaces connected to different VLANs, such the Sub LAN and Utility LAN shown in Fig.~\ref{fig:logical_CORE_network}, to emulate a wide-area-network (WAN) between substations and UCC. CORE also has a bridge connecting the cyber-physical EMS application (Section \ref{subsub:cypres}) that monitors real-time traffic from PWDS as well as network traffic in CORE. The WAN setup has direct connections between the gateway routers of the UCC and substation subnets. The routes within this architecture are created by running Quagga~\cite{quagga} services in the routers which employ open shortest path first (OSPF)~\cite{core_ospf} as the routing protocol. 

\begin{figure}[htbp]
%\centering
%\centerline{
\includegraphics[scale=.30]{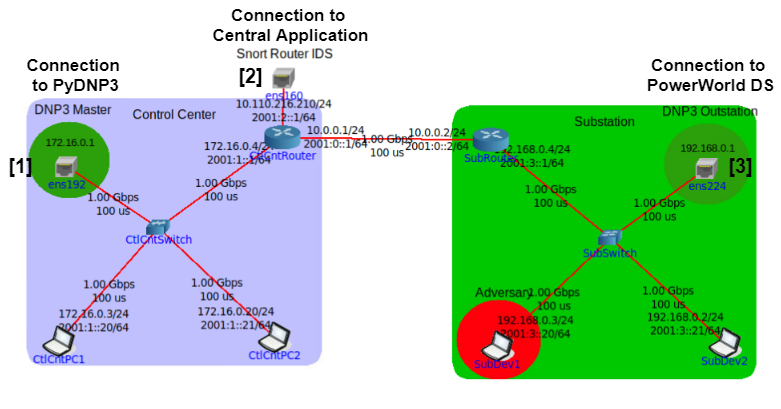}
%}
\caption{CORE network topology showing emulated PC nodes and connections to: [1] DNP3 master, [2] CYPRES app, and [3] PWDS DNP3 outstations. }
\label{fig:CORE_network}
\end{figure}

From left to right in Fig.~\ref{fig:logical_CORE_network}, the connections are: (1) VM hosting DNP3 master, (2) VM running CORE, (3) VM running a centralized cyber-physical EMS application, and (4) the PWDS VM. To show the emulated network, Fig.~\ref{fig:CORE_network} details the network topology:
the DNP3 master and SEL-RTAC are connected to the CORE through virtual interface [1]; interface [2] forwards Snort IDS alerts from control center router to the EMS application;
the VM running the large-scale synthetic electric case in PWDS is connected through 
interface [3].

\subsection{Power system simulation: PWDS}

In RESLab, PWDS 
%(Section ~\ref{subsec:virtualizationandemulation})
models the dynamic behavior of an electric power system in the transient stability time frame. It does interactive control~\cite{synthetic}, and serves as a general interface for DNP3 outstations~\cite{DS,overbye2017interactive}.  An outstation, as defined from a power system operational point of view, typically includes one substation and its devices, including branch breakers, generators, load breakers, and shunts. The DNP3 tags generate binary data, such as the status of all devices, and allow the devices to be controlled by other DNP3 masters/clients. The DNP3 tags can also be set to send analog data, such as measurements of generator real and reactive output, branch power flow, and bus voltage, as well as allowing DNP3 masters to change generator setpoints.  A DNP3 master is hosted in a VM running OpenDNP3. Another VM is running SEL acSELerator software~\cite{rtac_acselerator} and is used to configure the RTAC as a DNP3 master. 

PWDS also serves as a simulation engine with a generic interface for integration into other applications~\cite{w4ips}. In our experiments , PWDS simulates the power system in a real-time environment in which cyber threats and defense mechanisms are implemented.  The large-scale test case on the Texas footprint~\cite{synthetic, synthetic_comm} is implemented as our exemplar power system and is being maintained at Texas A\&M.
CORE’s WAN is being used to forward breaker status and control commands
between VMs.

\subsection{DNP3 and master application} 
DNP3 is extensively used by electric utility companies in North America for communication between equipment~\cite{dnp3_na}. %The protocol was released in 1993 suited to use the RS-485 serial protocol but has since been upgraded to work on TCP/IP networks. 
The protocol utilizes the master/outstation architecture.  A network can be configured to have one DNP3 master communicate with more than one DNP3 outstation, referred to as a multi-drop network. Alternatively, there can be one DNP3 master that communicates with one DNP3 outstation, or a one-on-one network, as implemented in this paper.

DNP3 messages contain a 10-octet DNP3 header and a maximum 292-octet DNP3 payload, which are carried over TCP/IP packets. The DNP3 header contains sync, length, link control, destination, and source address fields with a cyclic redundancy check (CRC) to ensure data integrity. The DNP3 payload is comprised of many 16-octet data blocks, with a 2-octet CRC for each block.

The purpose of the CRC is to ensure that bits have not been changed accidentally during its journey from source to end node. Some intruders may modify the traffic yet fail to modify the CRC which can be easily detected at the receiver or by implementing DNP3 specific decoders in IDS.
Inside the DNP3 payload, function codes identify the operation the outstation performs. The index identifies the device in the outstation the master is asking to perform the operation on or retrieve data from.
These are the function codes used in our simulations: Confirm (0x00), Read (0x01), Read (0x2), Select (0x03), Operate (0x04) Direct Operate with Acknowledge (0x05), Solicited Response (0x81), and Unsolicited Response (0x82).

The DNP3 master application in RESLab uses 
%a python program written by the authors that uses 
the PyDNP3 library, a Python wrapper for the C++ based OpenDNP3 module, to run the master as a console and a graphical user interface (GUI) application. The purpose of the master application is to continuously monitor the status of the circuit breakers, generators, and loads in the DNP3 outstations that are running in PWDS. The application also forwards the response of DNP3 outstations, as well as connection status, to the central application via CORE’s WAN. This application is configurable to change the polling rates and visualize real-time traffic. It runs in an isolated VM but exists in the UCC LAN with its default gateway set to 172.16.0.4, which is the UCC router (see Fig.~\ref{fig:CORE_network}).

\subsection{RTAC integration} 
RESLab also incorporates the SEL-3530 RTAC to explore different variants of DNP3 master. The RTAC provides flexible system control with integrated management of security, configuration, and logic. It supports multiple communication protocols, such as DNP3, Modbus, and IEC~61850, and comes with an embedded IEC~61131 logic engine~\cite{sel_2020}. RTAC has been utilized in several hardware-in-the-loop testbeds for data collection and signal conversion~\cite{kezunovic2017use, tamara_pnnl}, but they do not use it for communication studies or to emulate cyber adversaries associated with specific hardware.

Within RESLab, for each substation there is a DNP3 master in the RTAC to collect analog input data, such as power flow, current, and generator output, in addition to binary input data, such as the status of branches, generators, loads, and shunts from PWDS. Furthermore, each client in the RTAC can control the corresponding devices through analog and binary outputs such as to change generator setpoint in Mega Watts (MW) and device status (on/off). Thus, the integration of an industrial standard control device in RESLab allows researchers to gain a deeper understanding of how cyber adversaries can impact the devices and the system as well as develop more practical detection and defense logic in the field.

\subsection{Cyber-physical energy management system}\label{subsub:cypres}
A centralized cyber-physical energy management application that our team developed named Cyber-Physical Resilient Energy Systems (CYPRES) is designed to house algorithms for monitoring and analysis, run SCADA applications, and visualize the system.
CYPRES is developed and deployed in RESLab as an exemplar use case for the testbed. CYPRES aggregates information from the cyber side CORE emulation environment, the power side from PWDS, as well as from the DNP3 masters regarding DNP3 communication status. CYPRES is used to visualize the control network of the synthetic utilities and their substations in the synthetic power grid. To detect instrusions, it also probes real-time traffic, where CYPRES then performs data fusion from multiple sensors in the synthetic network. The CYPRES application is currently envisioned to be housed at a central location (i.e., at a balancing authority or utility) and used to analyze the system with respect to cyber intrusions.  Furthermore, CYPRES provides cyber-physical situational awareness in RESLab using attack tree visualizations. These visualizations can be tailored to network administrators or to power system operators to provide an actionable map of risk related to cyber and physical assets and impact, and to recommend mitigation actions for the identified risks within a network, e.g., by informing how to protect against cascading failures.
%%%%The details of different functionality of CYPRES application will go in another journal on CPS modeling %%%%%%%%%%%%%%% 

\subsection{Intrusion detection system} 
The role of an IDS is to detect cyber intrusions. Rule-based and anomaly-based IDS's are predominantly used in industry, but they lack the capability of detecting zero-day attacks. As an initial approach, RESLab integrates the Snort IDS which is used to detect and generate alerts for cyber intrusions. Snort tutorials~\cite{snort_cookbook} are followed to define rule sets, preprocessors, decoders 
and change configuration. 
Currently in RESLab, Snort is protecting the control center and substation LANs by running as a service within the routers. The alerts are forwarded to the CYPRES application in real-time.

For the synthetic power system case, the dataset from the attacks comprises packets with a destination port of 20,000, which is the default DNP3 port. 
From the filtered dataset, the frequency of communication between the master and outstation is analyzed in Section~\ref{results} to show to the effectiveness of the exemplar cyber intrusion scenario in impeding DNP3 communications.

\subsection{Storage and visualization}

RESLab implements a platform that the team has created to probe the traffic at all the network interfaces inside CORE, to collect the traffic, to use Elasticsearch Logstash Kibana (ELK) stack to store the traffic in an Elasticsearch index, and to visualize them using Kibana dashboard with the Packetbeat plugin~\cite{packetbeat}. One can configure the Packetbeat plugin to modify the number of interfaces and the type of traffic to probe. Kibana provides a platform to write Lucene queries to filter out a search in the Elasticsearch index. RESLab uses Logstash to collect Snort alerts to visualize in Kibana.  
In addition to ELK stack, RESLab also integrates Zabbix~\cite{zabbix} for network monitoring, as it provides a platform to configure custom alert rules and triggers. We have configured a Zabbix server in the base operating system hosting CORE, and the Zabbix agents in all the routers in CORE.
The agents within CORE use the CORE control network to interact with the server using ZBX protocol~\cite{zabbix}.

%%%%%%%%%%%%%Section 4%%%%%%%%%%%%%%%%%%%%%%%%%%%%%%%%%%%%

\section{Threat Model}~\label{cyber_threats}

Widely-known cyber threats such as the Ukranian and Stuxnet attacks 
have been multi-stage attacks, 
which are a serious concern. However, due to the nature of these attacks, and the wide range of time scales involved at each stage, they are challenging to plan 
and study.
Hence, RESLab is intended to mimic an electrical utility environment allowing for experimentation of individual threats, which enables us to develop and test solutions at each step in the attack vector. %of such an attack.

The threat model we present and implement in this paper is based on emulating a 
multi-stage attack
in the large-scale synthetic test system’s communication model. In the first stage, the adversary gains Secure Shell (SSH) access to %one of the host 
a  machine in the substation LAN. In the second stage, the adversary performs steps that are tailored to the system under study and to power system protocols, allowing the adversary to achieve 
MiTM and DoS attacks that cause physical impact.
The RESLab framework can not only support MiTM and DOS, but can also integrate other attack vectors.

\subsection{Man-in-The-Middle (MiTM) attack}
MiTM is one of the oldest forms of cyber intrusion, where a perpetrator positions itself in a conversation between two end points, to either passively eavesdrop or to impersonate one of the endpoints, making it appear to be a normal exchange of information. MiTM encompasses different techniques and potential outcomes, depending on the threat model. During the second stage of our presented threat model, we compromise the target outstation and its router through performing an ARP spoof attack by poisoning the ARP cache of both the substation's gateway and DNP3 outstation~\cite{arp_cache_poison}. Then, in the third stage, we modify control and monitoring traffic to have different implications on the electric grid.

Such tampering of commands and measurements would normally go undetected by the outstation using 
CRC error checking, since the data chunk in the DNP3 payload’s has its CRC recalculated by the adversary before the modified packet is forwarded to the outstation. The intruder causes false command injection (FCI) and false data injection (FDI) attacks by first storing the DNP3 polling response for the targeted outstations, then manipulating measurements in some cases, and commands in other cases, as well as manipulating a mix of both to carry out one of the most critical contingencies presented in our N-x contingency discovery paper~\cite{n_x}. Such an attack is hard to be detected by an IDS such as Snort,
if the intruder not only tampers the command and takes care of the CRCs.

In RESLab, the MiTM attack is developed and implemented to change binary and analog commands sent by the DNP3 master to the outstation as well as the polled response from DNP3 outstations. The intruder not only modifies commands but also eavesdrops and then modifies the current state of the system by tampering its real-time measurements. In Table ~\ref{table:mitm_attack_procedure}, the procedure for performing a MiTM attack in RESLab is listed. 
The details on the various combination of attacks that are performed in the third stage of the threat model is presented through four use cases detailed in Section~\ref{use_cases}.

\begin{table}[b]

%\begin{tabular}{|c{1.0cm}|m{7.2cm}|}
\begin{tabular}{|p{1.0cm}|p{7.2cm}|}
\cline{1-2}
 Seq. & Description   \\ \cline{1-2}
 1 &  Start the CORE, PWDS, OpenDNP3 master.Allow time for DNP3 communication between master and outstation to be established. \\ \cline{1-2}
 2 &  Start CYPRES app. to monitor cyber data.Start running Snort in substation router. Run the ELK services and Packetbeat .\\
      \cline{1-2}
 3 &  ARP cache poisoning of substation's gateway and outstation. \\
     \cline{1-2}
 4 &  Sniff traffic to and from the outstation.Forward non DNP3 traffic to/from outstation.\\
  \cline{1-2}
 5 & Send command from master to outstation. Modify command and forward to outstation.
      \\ \cline{1-2}
 6 &  Modify TCP acknowledgement (ACK) from outstation.
         \\ \cline{1-2}
\end{tabular}
\caption{The steps taken in RESLab to implement FCI injection.}
\label{table:mitm_attack_procedure}
\vspace{-0.5cm}
\end{table}

\subsection{Denial of Service (DoS) attack}
As another attack vector, we implement a DoS attack
to exhaust victim nodes’ processing capability and link bandwidth. 
There are many different methods of DoS attacks that can be used, which include but are not limited to: user datagram protocol (UDP) flood, internet control message protocol (ICMP) flood, and Ping of Death (PoD)~\cite{dos_scada}. While each of these DoS attack types use different Open Systems Interconnection (OSI) 
layers such as application, presentation, session, transport,
network, data link or even at physical layer protocols
to carry out the attack, all DoS methods attempt to disrupt the communication channels of the targeted node. In our threat model, the intruder within the substation LAN targets routers at the substation and at the control center
by flooding the routers with ICMP traffic. 
The impact of these DoS attacks is then observed and analyzed based on round trip times (RTT) and throughput of the communication channel by varying the strength of the attacks such as the length and delay between the ICMP packets infused to disrupt the DNP3 session.

%%%%%%%%%%%%%%%%%%%%%%%%Section 5%%%%%%%%%%%%%%%%%%%%%%%%%%%%%%%%%%%%%%%%%
\section{Impact of Cyber Threats on Power Operation}
\label{use_cases}
  
\begin{figure}[t!]
\centerline{\includegraphics[height=2.3 in,width=3.5 in]{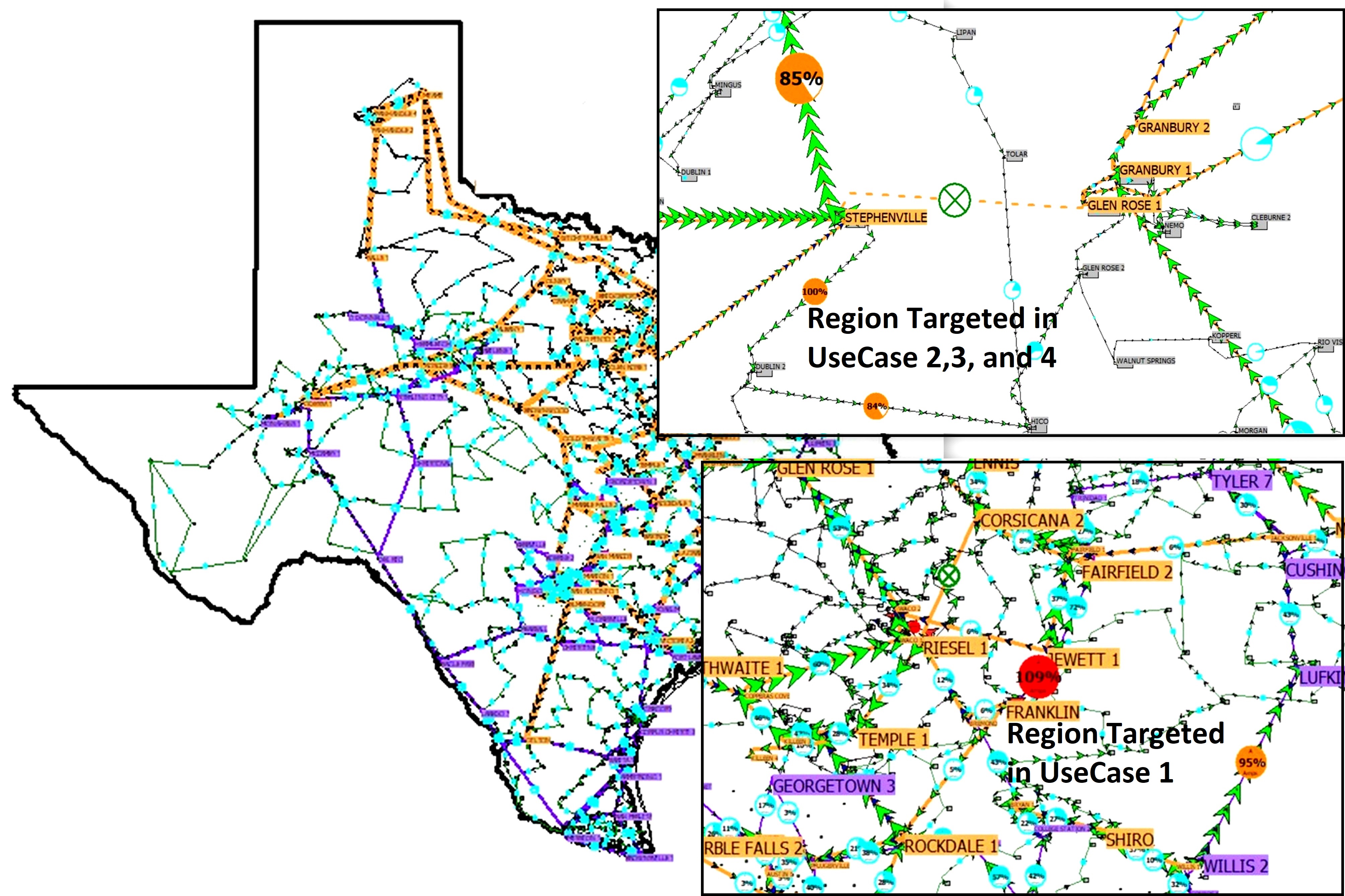}}
\caption{The Texas-2000 bus model is the basis of our exemplar cyber-physical power system.
}
\label{fig:phy_impacts}
\end{figure}

The synthetic Texas 2000-bus case~\cite{synthetic, synthetic_comm} is a publicly available power system test case. This system is N-1 secure, which means the system can still operate securely with one device is out, and it is difficult to cause disruption by exploring N-2 contingencies. Hence, the use cases in RESLab leverage results from our prior work~\cite{n_x} on identifying the most critical multiple-element contingencies based on graph theory and line outage distribution factors (LODFs), which are located in the regions targeted in our use cases (Fig.~\ref{fig:phy_impacts}). 

Assume branch (x,y) means from Bus x to Bus y in the Texas 2000-bus model. To illustrate the contingencies, if branches (5262,5260), (5263,5260), (5317,5260), (5358,5179) are open, there will be four overflow branches in the system, which are branches (5071,5359), (5138,5071), (8086,8083), (8084,8083). Branch (5262,5260) and (5263,5260) are located in Substation GLEN ROSE 1 (560), Branch (5317, 5260) is at Substation GRANBURY 1 (601), and Branch (5358, 5260) is at Substation RIESEL 1 (631). The overflow branches are at Substation WACO3 (399), JEWETT1 (1195) and FRANKLIN (1200).

Besides, in those substations there are several generators. We have studied that if those generators are compromised, there will also be a contingency in the system.  These generators are Gen 5262, 5263, 5319, 5321, 5360, 7098, 7099. Gen 5262 and 5263 are at Substation GLEN ROSE 1 (399), Gen 5319 and 5321 are at Substation GRANBURY 1 (601), Gen 6360 is at Substation RIESEL 1 (631), and Gen 7098 and 7099 are at Substation WADSWORTH (968). When these generators reduce their output and the Branch (5260,5045) in Substation STEPHENVILLE (390) is open, there will be another overflow in Branch (5286, 5046) at Substation STEPHENVILLE (390).
 
Thus we assume that the adversary 
 %is motivated,
has the intent and the resources to target the most critical branches and generators, where disrupting their control causes severe impact.  Specifically, we present four use cases to show how cyber threats can compromise a resilient power system.
 %from its underlying network. 
These use cases involve binary and analog command modification, measurement and status modification.
Before exploring the 
%various 
scenarios, we present RESLab's experimental setup 
%for these scenarios 
%in
%of
%RESLab, 
which allows us to collect data at various locations and analyze them from a cyber-physical perspective.

\subsection{Experimental setup}

\label{experiment}
The DoS and the MiTM attacks for the scenarios are performed while running RTAC and OpenDNP3 applications as the DNP3 masters. The resources used to perform all the experiments are illustrated in Table~\ref{table:resources}. Virtual LANs (VLANs) are used to ensure that traffic is forwarded by the emulated routers in CORE and to segregate the substation network from the control center network. 
%Without the VLANs, network DNP3 traffic could circumvent CORE’s emulated network and go through vSphere’s tap interface connected to the internet, or other VLANs.

\begin{table}[b]
\begin{center}
\begin{tabular}{ |p{1.95cm}|p{0.75cm}|p{0.75cm}|p{1.2cm}|p{1.9cm}|  }
 \hline
 \multicolumn{5}{|c|}{Virtual machine allocations in vSphere} \\
 \hline
 VM Name & Mem. & CPU Cores & VLANs & OS\\
 \hline
 CORE  & 12G    & 4 &   1,2,3,4 & Ubuntu\\
  \hline
 DNP3\_Master &   12G  & 4   & 1,2,3 & Ubuntu\\
  \hline
 PWDS & 10G & 2 & 1,2,4 & Windows 10\\
  \hline
 Central\_App & 16G & 8 & 1,2 & Windows 10\\
 \hline
 RTAC & 4G & 2 & 1,2,3 & Windows 10\\
  \hline
\end{tabular}
 \caption{VM configuration for the RESLab architecture in Fig.~\ref{fig:logical_CORE_network}. }
\label{table:resources}
\end{center}
\vspace{-0.5cm}
\end{table}

In these simulations, we use a multi-master architecture where each master monitors and controls a substation separately. While the master monitors and controls outstations,
the adversary sniffs all the measurement traffic (requests and responses) from the substations.
We capture network traffic at four locations in the network (outstation, master, adversary and substation router) to evaluate the impact of MiTM attacks on these four use cases. Since the adversary acts as the middle man between the substation router and outstation, we validate the MiTM by checking if the DNP3 packets received at the substation router and at the master are identical, and if the DNP3 packets at the outsation and adversary are identical. 
To test the detection of DoS and MiTM attacks,
we operate Snort at the substation and control center routers in a Network Intrusion Detection System (NIDS) mode by enabling preprocessors and decoders and including custom rules for ARP, DNP3, and ICMP traffic. Then, we present the alerts along with the physical traffic to correlate the alerts with the measurements and command tampering.
%} 

\subsection{Class 1: False Command Injection (FCI)}

\begin{figure}[t!]
\centerline{\includegraphics[height=2.3 in,width=3.5 in]{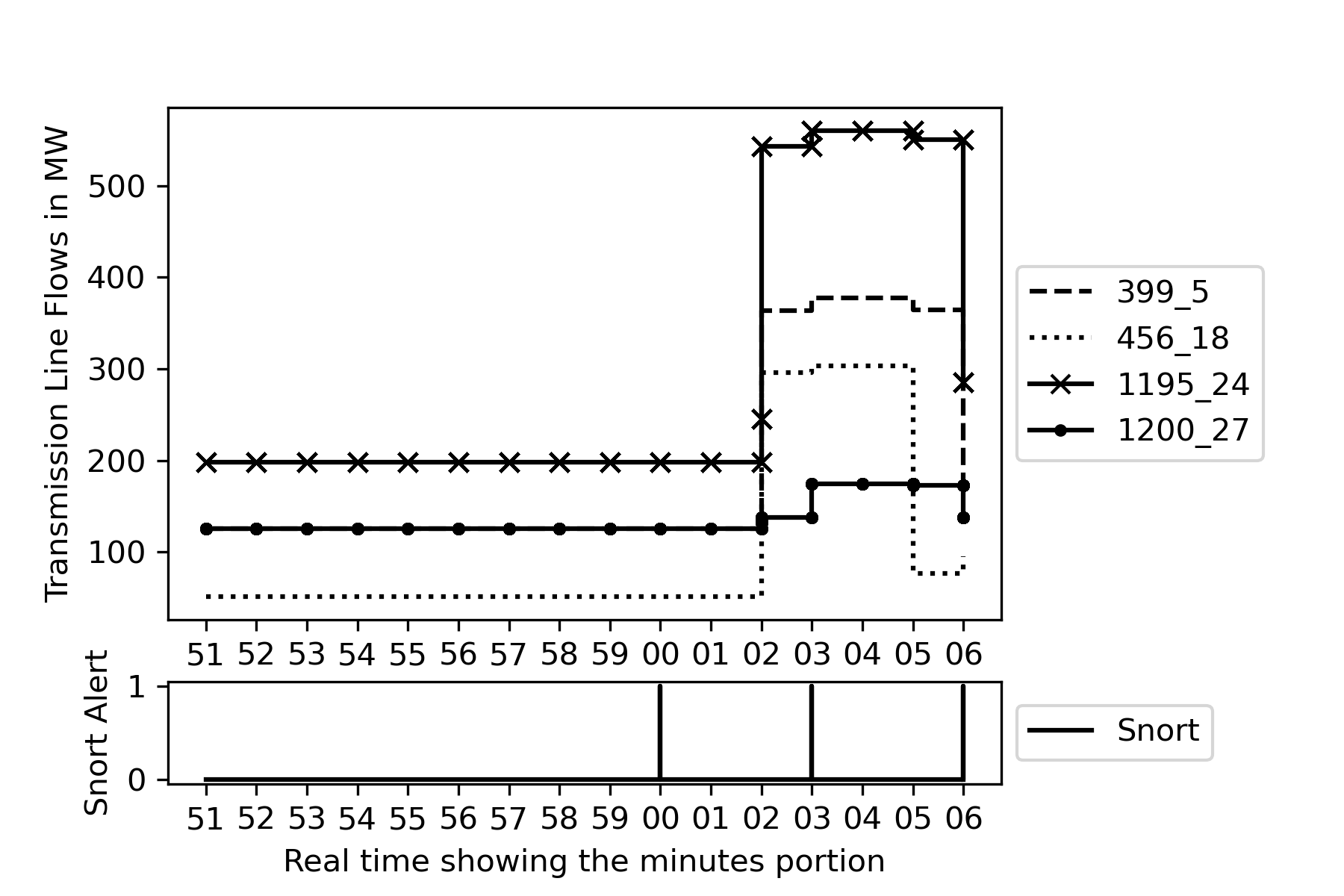}}
\caption{\textit{\textbf{UC1:}} Overloaded transmission lines observed at master application  (\textit{WACO 3} (399), \textit{WACO 1} (456), \textit{JEWETT 1} (1195), and \textit{FRANKLIN} (1200)). The legend shows the $outstation\_index$, for example, the first legend indicates outstation number 399 and DNP3 index 5. The plot beneath shows the Snort alerts during the intrusion.
}
\label{fig:uc1_line_of}
\end{figure}

%For instance, 
A MiTM attack that modifies binary control commands using relay control blocks can cause line overloading~\cite{dnp3_crob}. To achieve this, the adversary first parses the measurements by sniffing the DNP3 responses from the outstation. Then, it sniffs the DNP3 binary \textit{OPERATE} command and %traffic associated with the identified Outstations (substations) listed above and 
forges them.
%binary commands. 
The adversary modifies commands with function codes of 3 and 4 (\textit{SELECT} and \textit{OPERATE} command) from the RTAC, and it modifies commands with function code 5 (\textit{DIRECT OPERATE} command) from the OpenDNP3 master application. 
%Then, 
The adversary modifies all the \textit{CLOSE} commands to \textit{TRIP}, forcing to open the critical branches identified and causing line overloads in four other branches, shown in the data from the scenario in Fig.~\ref{fig:uc1_line_of}. This scenario is referred as \textit{use case 1 (UC1)}.

The intruder can also modify analog control commands to change the setpoints in generators along with a binary command to control a branch to cause line overloads. The intruder first inspects the DNP3 packets, 
changes a collection of generator setpoints from the real value to 0, and %also 
alters the binary control command as in \textit{UC1}. This scenario compromises seven generators and one branch, referred as \textit{use case 2 (UC2)}.

\begin{figure}[t!]
\centerline{\includegraphics[height=2.3 in,width=3.3 in]{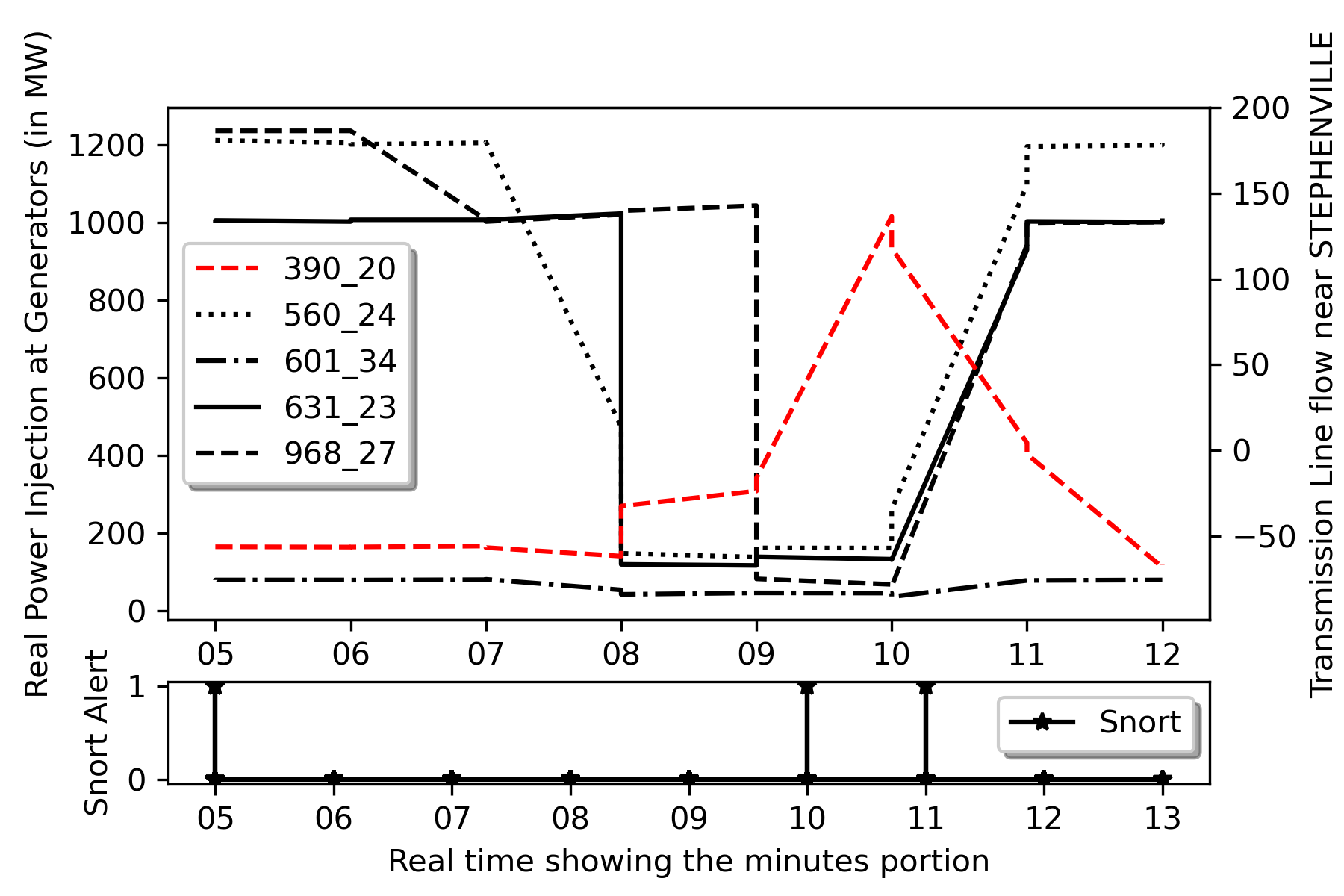}}
\caption{\textit{\textbf{UC2:}} The real power injection at generators from substations \textit{WADSWORTH} (968), \textit{RIESEL 1} (631), \textit{GRANBURY 1} (601), \textit{GLEN ROSE 1} (560) (left y-axis) and the overloaded line near substation \textit{STEPHENVILLE} (390) (right y-axis). The legend shows the $outstation\_index$.
The plot beneath shows the Snort alerts.
}
\label{fig:uc2_phy_2}
\end{figure}

Fig.~\ref{fig:uc2_phy_2} shows the actual generation output in each substation \textit{WADSWORTH}, \textit{RIESEL}, \textit{GRANBURY}, and \textit{GLEN ROSE} along with the Snort alerts during
5th, 10th, and 11th
% $5_{th}$, $10_{th}$, and $11_{th}$ 
mins of the scenario. The intrusion in these substations takes place during the 
8th and 9th
% $8_{th}$ and $9_{th}$
mins. The intruder’s goal is to overload the transmission line near substation \textit{STEPHENVILLE}, %(outstation no. 390), 
%which is 
accomplished by tampering of the analog set points, as observed in the interval 9-11th minute in Fig.~\ref{fig:uc2_phy_2}.

\subsection{Class 2: False Data Injection (FDI) with FCI}

The MiTM intruder can also perform FDI with the FCI to create more difficult-to-detect attacks.
First, the intruder falsifies polled measurements, causing the operator to re-send a control command to the field device. Then, the intruder modifies the control command, as in the previous use cases, by changing the generator setpoint.  The actual generation measurements 
for the same seven generators in \textit{UC2}
are falsified to 20~MW, and the flow measurement coming from branch [5260, 5045] is changed to 3000~MW, which is above its capacity. Based on these observations, the operators or a pre-defined control logic within devices such as an SEL RTAC, would re-send the control command to increase the generators’ output and open the branch. However, once sending those commands, the intruder modifies the setpoints to 20~MW, making the physical system unreliable. This scenario is referred as \textit{use case (UC3)}.

\begin{figure}[t!]
\centerline{\includegraphics[height=2.3 in,width=3.5 in]{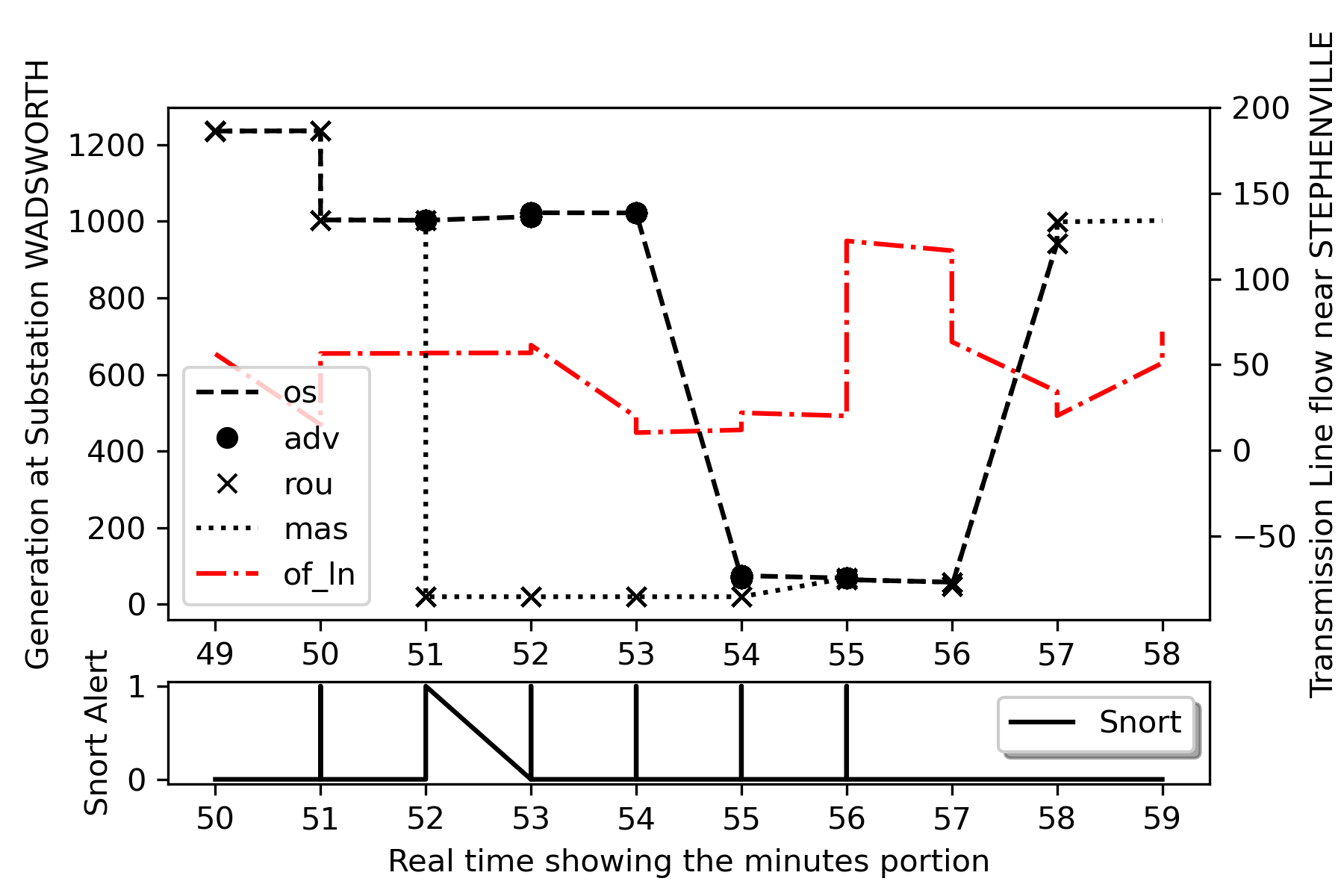}}
\caption{\textit{\textbf{UC3:}} The real power injection at one generator in substation \textit{WADSWORTH} as observed by master (\textit{mas}), substation router (\textit{rou}), adversary (\textit{adv}), and the outstation (\textit{os}), along with the overloaded line (\textit{of\_ln}) near substation \textit{STEPHENVILLE}. The plot beneath shows the Snort alerts during the intrusion.}
\label{fig:uc3_phy}
\end{figure}

Fig.~\ref{fig:uc3_phy} shows the system after the output of a generator in substation \textit{WADSWORTH}
is changed in the polled measurements by the intruder from 1000 MW to 20 MW
as observed in the master and the router during 52nd to 55th min. 
The Snort alerts are observed from the 53rd to 56th mins. The alerts at 50th and 51st min are due to an attack in other targeted substations such as \textit{GLEN ROSE}, \textit{RIESEL}, \textit{GRANBURY}, whose generation set points are tampered. %\textcolor{red}{[clarify what is happening in other substation?]}.

Another example of a three-stage attack is referred as \textit{use case 4 (UC4)}, where the intruder first changes the measurements polled by the DNP3 master, as in \textit{UC3}. Once the operator re-sends the control command, the intruder changes the setpoints from the real value to a low value, as in \textit{UC2}, but the intruder also falsifies the measurement packets, masking the true measurements but showing the original setpoint values. The result is that the operator believes his/her command has been successfully received and committed. However, in the true physical system, the generators’ outputs are decreasing, and opening a line will then cause an overload.

\begin{figure}[t!]
\centerline{\includegraphics[height=2.3 in,width=3.5 in]{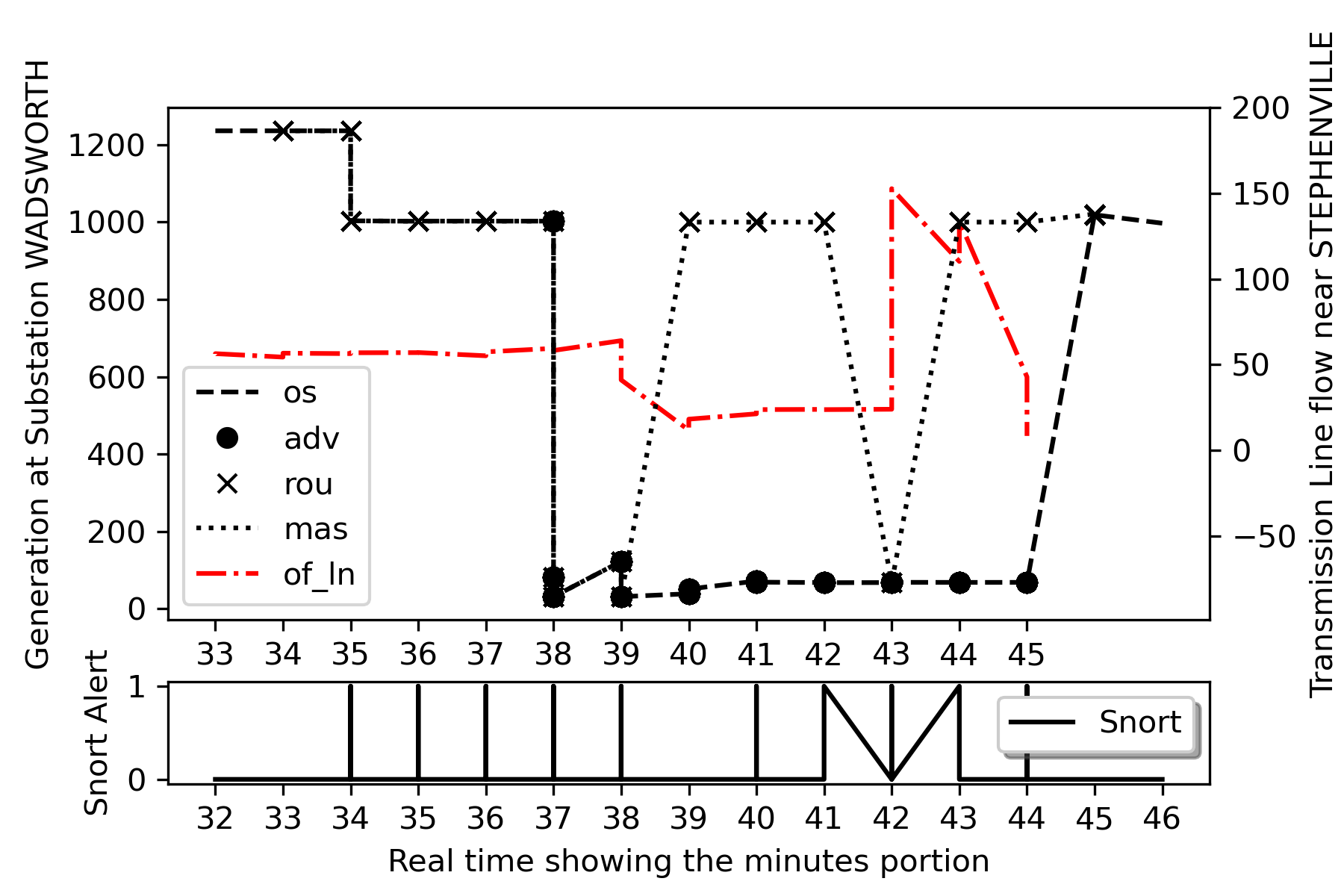}}
\caption{\textit{\textbf{UC4:}} The real power injection at one generator in substation \textit{WADSWORTH} as observed by master(\textit{mas}), router(\textit{rou}), adversary(\textit{adv}), and the outstation(\textit{os}). The overloaded line magnitude (right y-axis) near \textit{STEPHENVILLE} (\textit{of\_ln}).  }
\label{fig:uc4_phy}
\end{figure}

Fig.~\ref{fig:uc4_phy} shows the generation output at substation \textit{WADSWORTH} as observed at four locations. During the intrusion on \textit{WADSWORTH}, within the 34th and 44th mins, the adversary first forces the master to take a wrong action to change the generation output to 1000~MW once the master observes low generation output at the 34th minute due to modification of the measurements of generation output. 
Further, when the operator takes this action to address the low generation output,
%Further,
the intruder changes the command from 1000~MW to 0~MW to cause contingency.
To be stealthier, the intruder also modifies the polled response from outstation with the same setpoint value of 1000~MW from the interval 39 to 44 min except at 42nd min, as set by the operator, to prevent the master from observing the contingency caused by the intruder in first two stages. The snort alerts generated in this interval are shown in Fig.~\ref{fig:uc4_phy}.

%%%%%%%%%%%%%%Section 6%%%%%%%%%%%%%%%%%%%%%%%%%%%%%%%

\section{Results and Analysis}
\label{results}

In this section, DoS and MiTM attacks are performed and analyzed on the DNP3 sessions between the masters and outstations based on the four use cases, which are summarized in Table~\ref{tab:uc_table}. The effectiveness of the DoS attack is evaluated by varying attack strength and studying its impact on the %Round Trip Times (
RTT and throughput of DNP3 traffic. The time frame of power system operations compared with the attack time frame plays a major role; for example, the time frame of inverter and stator transient control are in the order of milliseconds, while control of voltage stability, power flow, and unit dispatch %\textcolor{red}{I don’t think you meant to say unit commitment here} \as{I wanted to mention about the variability of interval of different operation} 
range from 10 to 1000 seconds. Hence, it is essential to minimize RTT to ensure that the control commands are processed by the field devices on time. 
%impact creating higher RTT is essential to mitigate 

The use cases for MiTM attacks are tested with the RTAC and OpenDNP3 master. Experiments are conducted by varying the number of DNP3 masters, as well as the polling interval. Each master communicates with its substation, and we assume that 5 or 10 master are connected with their respective outstations. 
%(i.e., 5 or 10 substations). 
These experiments are performed to study the success rates of the attacker in causing the desired contingency of each use case. The adversary is restricted by the available resources in the Linux containers in CORE, thus the attacks are stochastic in nature. As the number of masters increases, the amount of traffic an intruder processes increases, which results in higher attack miss rates, i.e., the probability that the attacker fails to modify a sniffed packet. The results demonstrate the effective implementation of the use cases by observing the real-time physical side data at different locations in the testbed. 

We also study the number of active TCP connections as impacted by retransmission during the progression of the attack based on different polling rates and varying 
%the number of 
DNP3 masters. The adversary success probability, the average retransmission rates, the packet processing times, and the average RTT for performing each FCI and FDI attack, and the Snort alert statistics are key characteristics for detection.
Snort IDS is used to detect the ICMP flood 
attack as well as the ARP spoof attack (Section~\ref{cyber_threats}) that reroutes packets to the adversary and allows modifications to take place.

\subsection{DoS attack evaluation}

The DoS attack is performed by increasing the ICMP 
payload size for a fixed interval rate, as well as by varying the interval rate with  the fixed payload size, to determine their impact on the RTT and throughput of DNP3 traffic.  For all the experiments, the virtual Ethernet communication links in the network have a fixed bandwidth of 10~Mbps, with a 160~$\mu s$ transmission delay.

We first run polling at an interval of 30 and 60~s and \textit{DIRECT OPERATE}
commands
from the DNP3 master without any attack, to verify the network performance. Then, 
%for each proceeding trial of the attack, 
either the delay interval or payload length of DoS traffic is altered, and the same polling and control operations are performed.  

\begin{figure}[t]
\centerline{\includegraphics[height=2.0 in,width=3.5 in]{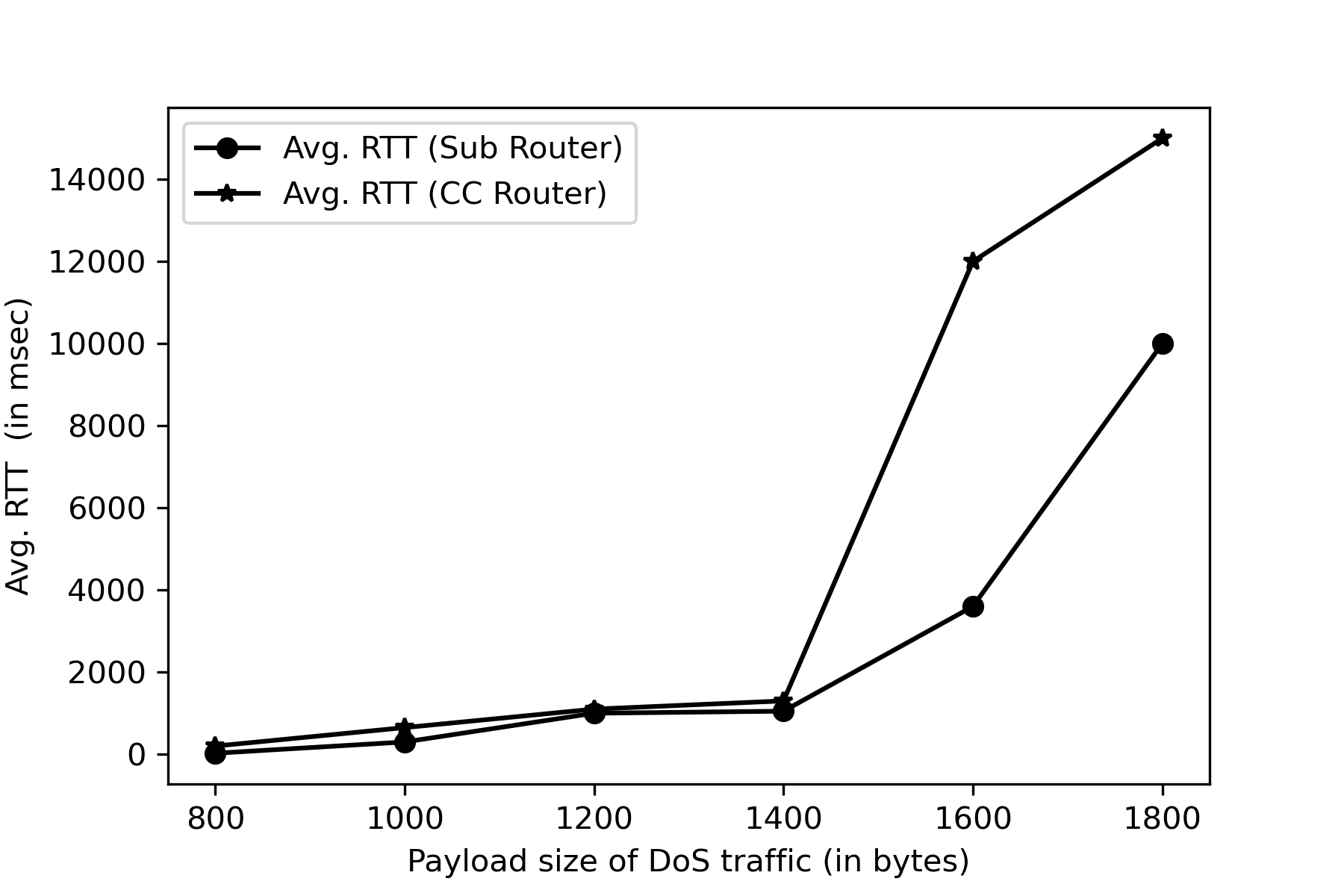}}
\caption{Impact of DoS on RTT by varying payload size.} %on RTT at both substation and control center routers.}
\label{fig:dos_rtt_payload}
\end{figure}

The DoS attack is performed by the compromised device in the substation, encircled in red, as shown in Fig.~\ref{fig:CORE_network}. The DoS attack is directed at two targets: the substation router and the control center router, and we seek to determine which device is impacted more by a DoS event.  From the DoS trials directed at the substation router, we observe that only the broadcast domain of the substation LAN is exhausted. However, if the attack is directed at the control center router, the link between the two routers as well as the broadcast domain of the substation LAN are exhausted.  
Both DoS attacks are performed using $hping3$~\cite{hping3} while keeping a fixed interval rate of 1000 ms and increasing the payload size of the ICMP packets from 800 bytes to 1800 bytes in increments of 200 bytes for each trial.  It can be observed from Fig.~\ref{fig:dos_rtt_payload} that average RTT increases with payload size, and that the attack on the control center router has a higher impact in comparison to substation router.

\begin{figure}[t]
\centerline{\includegraphics[height=2.0 in,width=3.5 in]{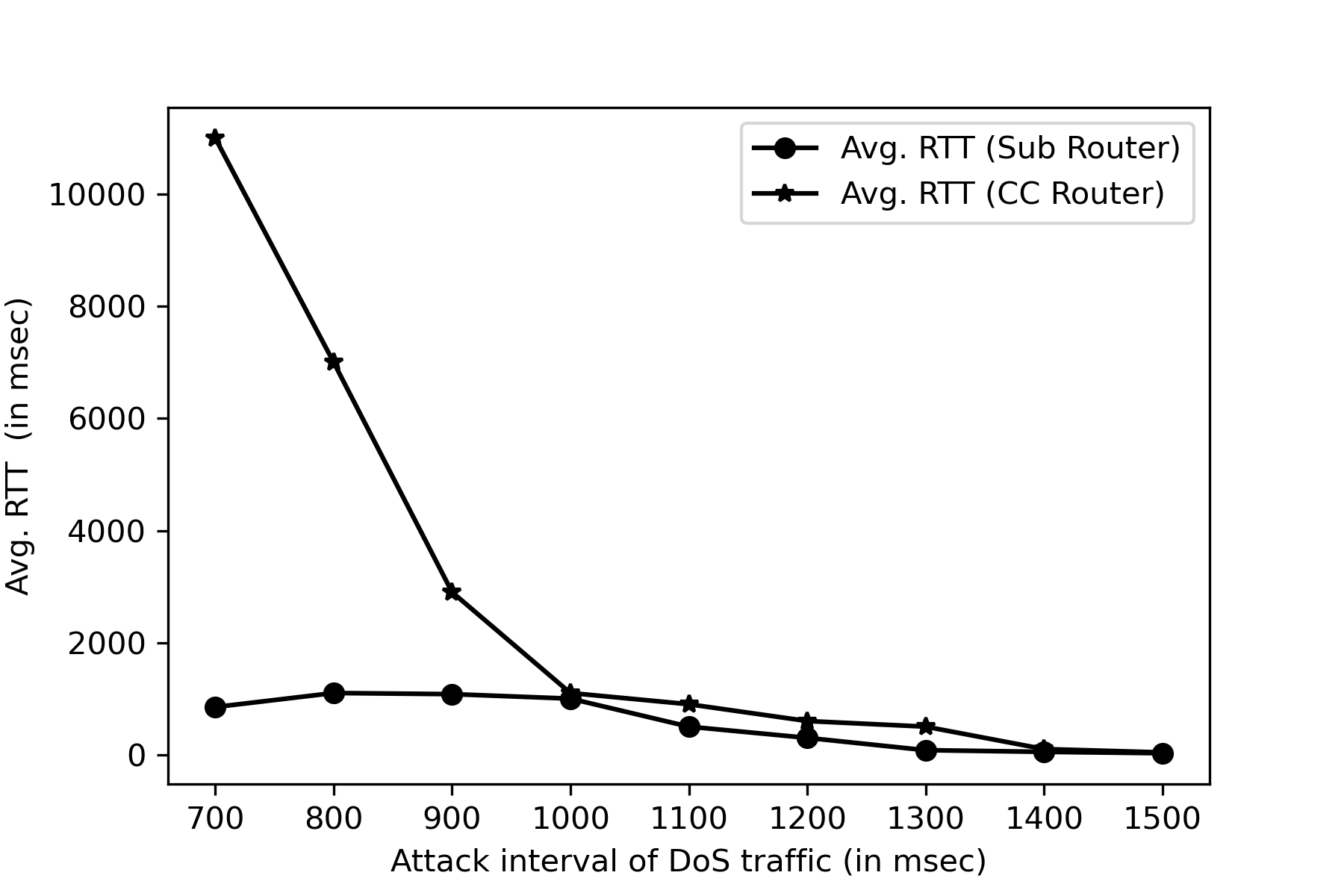}}
\caption{Impact of DoS on RTT by varying attack interval.} % on RTT at substation and control center routers.}
\label{fig:dos_rtt_interval}
\end{figure}

Further, DoS attacks are performed by keeping the payload size of the ICMP packet fixed at 1000~bytes while decreasing the ICMP packet arrival rate from 1500~ms to 500~ms in step decrements of 100~ms for each consecutive trial. 
%The targets of the attack are the gateways of both the control center and substation networks. 
Fig.~\ref{fig:dos_rtt_interval} shows that the average RTT decreases with increase in attack interval as well as the lower attack interval has higher impact on the control center router in comparison to the substation router.

%\begin{figure}[h!]

A DoS attack primarily affects the target’s downstream bandwidth. Hence, the average throughput will be affected as the bandwidth of the link is affected. The average throughput for the substation router is calculated using the transmission time of DNP3 packets, as per Eq.~\ref{eq:tp}:
\begin{equation}\label{eq:tp}
    Average\;Throughput = \frac{Total\;data\;payload\;in\; bytes}{Total \; transmission \; time}
\end{equation}

The average throughput depends on the command type from the DNP3 master. For example, the response payload size for the polling will be quite high compared to the response of the \textit{OPERATE} commands.  The goodput is equal to the throughput if there are no retransmissions.

\begin{figure}[t]
\centerline{\includegraphics
%[width=1.0\linewidth]
[height=2.0 in,width=3.5 in]
{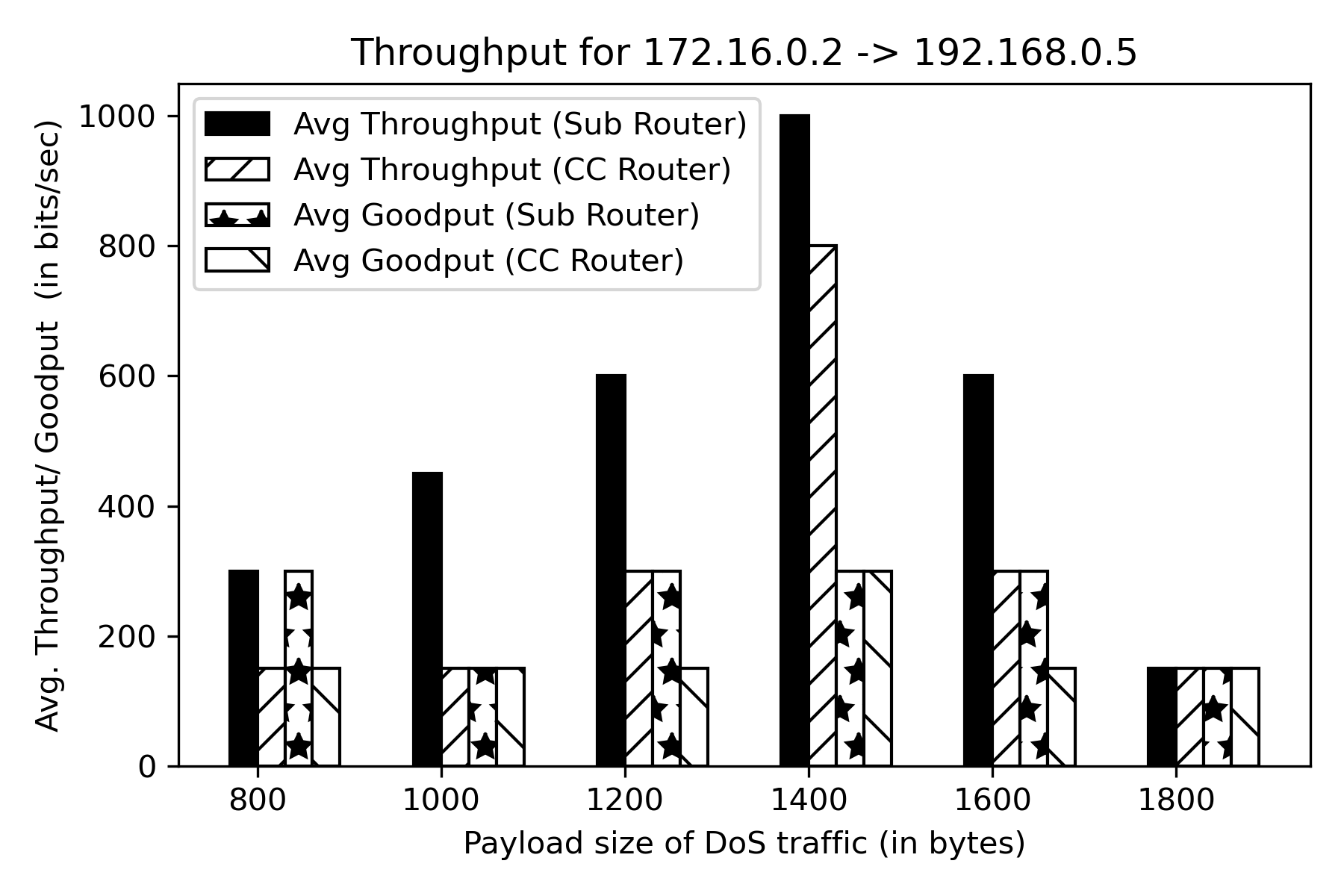}}
\caption{Impact of DoS on varying payload size on average throughput and goodput at both substation and control center routers.}
\label{fig:dos_tp_payload}
\end{figure}

\begin{figure}[t!]
\centerline{\includegraphics
%[width=1.0\linewidth]
[height=2.0 in,width=3.5 in]
{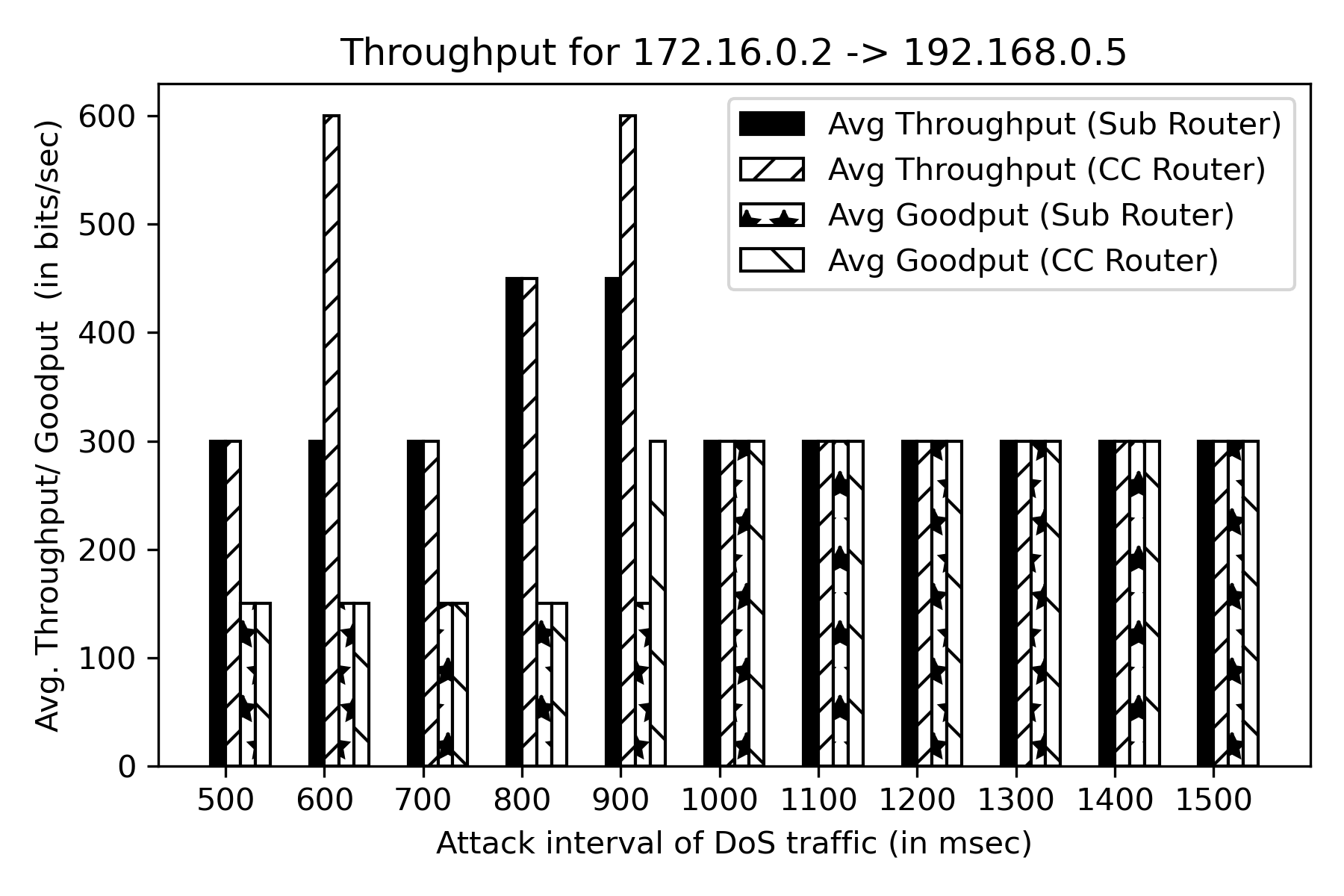}}
\caption{Impact of DoS on varying attack interval on average throughput and goodput at both substation and control center routers.}
\label{fig:dos_tp_interval}
\end{figure}

In Fig.~\ref{fig:dos_tp_payload}, we observe that the throughput and goodput increase as the payload size increases up to certain extent, then they decrease due to congestion in the network. It can also be observed that the difference between throughput and goodput increases as the payload size increases due to high retransmission caused by the congestion. 
Similarly, reduced goodput is also observed when the attack interval is lowered from 1500 ms to 500 ms as seen in Fig.~\ref{fig:dos_tp_interval}.

%\begin{figure}[h!]

\subsection{MiTM attack evaluation}
% perform mitm attack
% show the picture of the wireshark at the aatacker
% show the RTT at the DNP3 master
% show the DS figure dnp3 log check the close open status

\begin{figure}[t!]
\centerline{\includegraphics
[width=1.0\linewidth]
%[height=2.0 in,width=3.5 in]
{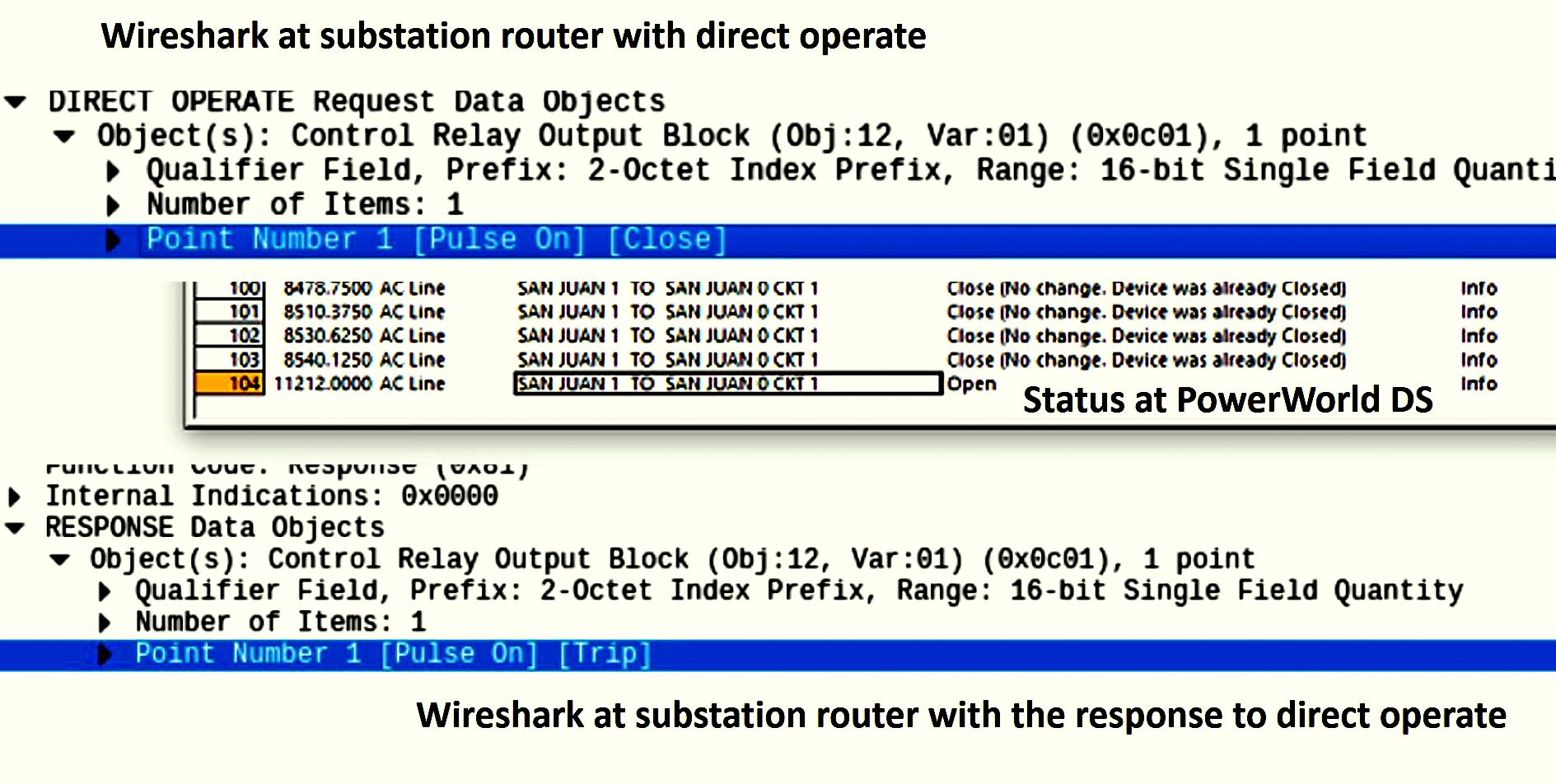}}
\caption{The DNP3 \textit{DIRECT OPERATE} command altered by the intruder.}
\label{fig:mitm_command_change}
\end{figure}

In the MiTM attacks, both master and outstation DNP3 packets are captured at the adversary’s machine located in substation LAN. Fig.~\ref{fig:mitm_command_change} shows Wireshark sniffing the DNP3 \textit{DIRECT OPERATE} command from the master in addition to the response from outstation.  As described in Section~\ref{use_cases}, the \textit{CLOSE} command is replaced by the \textit{TRIP} command as observed from the response, as well as the DNP3 log of PWDS as seen in Fig.~\ref{fig:mitm_command_change}.

The RTT for MiTM attacks is small compared to the RTT for a DoS attack. In a DoS attack, the RTT depends on the number of retransmissions, but in a MiTM attack, the RTT depends on how much time the attacker takes to parse the packet, modify the payload, recalculate the checksum and CRCs, and to forward the packet to the target. There is no substantial retransmission in the case of MiTM attacks, if the intrusion is stealthy.

\begin{figure}[t!]
\centerline{\includegraphics
%[width=1.0\linewidth]
[height=1.8 in,width=3.5 in]
{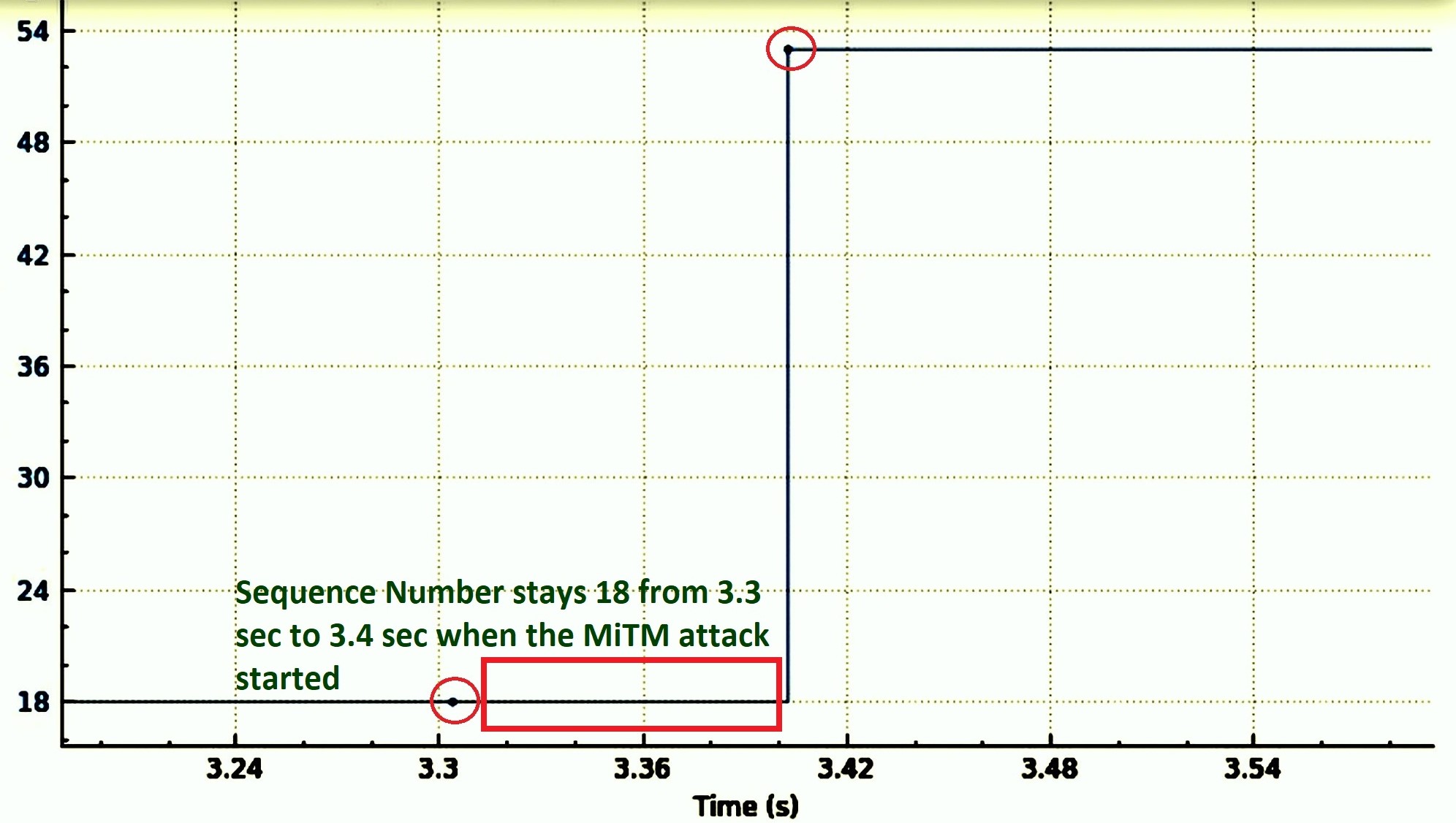}}
\caption{Verification that intruder used the same sequence number to forward modified packet.}
\label{fig:seq_attack}
\end{figure}

\begin{figure}[t!]
\centerline{\includegraphics[width=1.0\linewidth]{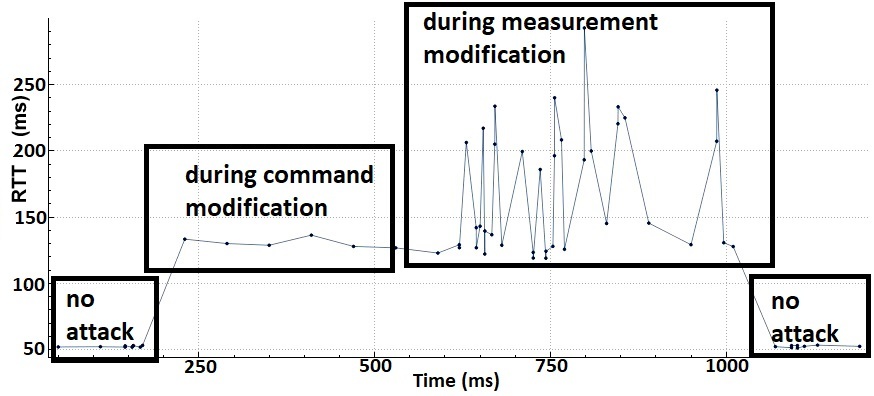}}
\caption{RTT of DNP3 traffic through the Substation Router during the FCI and FDI attack in use case 4 at substation \textit{WADSWORTH}.}
\label{fig:rtt_mitm_subRouter}
\end{figure}

The occurrence of a MiTM attack is validated both by observing a rise in RTT compared to the normal operation in Fig.~\ref{fig:rtt_mitm_subRouter} and from its
sequence number graph Fig.~\ref{fig:seq_attack}.
Specifically, in Fig.~\ref{fig:rtt_mitm_subRouter}, the MiTM attack is performed from 200~s to 1000~s, and the RTT is observed to increase to almost 150~ms during sniffing and FCI attack and to almost 200~ms during FDI attacks on measurement, indicating the time taken by the adversary’s machine for parsing and modification affects the overall RTT.  
Additionally, as the sequence number remains at 18 from 3.3~s to 3.4~s in Fig.~\ref{fig:seq_attack}, it indicates the attacker used the same sequence number to forward the modified packet. 

%evaluating MiTM attack based on the sequence number

\begin{figure*}[h]
\centerline{\includegraphics[width=1.0\linewidth]{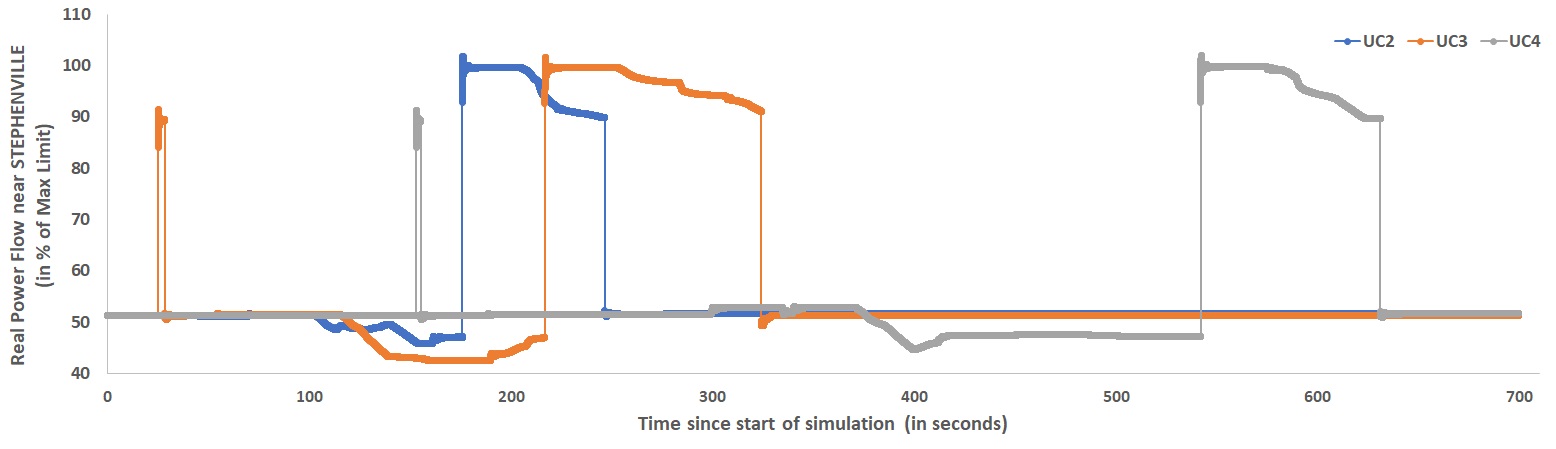}}
\caption{Impact of line overload caused through different use cases.}
\label{fig:flow_impacts}
\end{figure*}

\subsection{Use case specific physical impact evaluation}

\begin{table*}[t]
    \centering
    \begin{tabular}{|l|l|l|l|}
     \hline
 \multicolumn{2}{|c|}{FCI} & \multicolumn{2}{|c|}{FCI with FDI}\\
 %\hline
    \hline
         UC1 & UC2 & UC3 & UC4\\ \hline
        Bin. Commands & Alg.,Bin. Commands & Measurements=$>$Commands & Measurements=$>$ Commands=$>$Measurements \\ \hline
    \end{tabular}
    \caption{Use cases based on the type and sequence of modifications performed to study physical impacts.}
    \label{tab:uc_table}
    \vspace{-0.2 cm}
\end{table*}

The physical impact is evaluated based on the four use cases shown in Table~\ref{tab:uc_table}, described in detail in Section~\ref{use_cases}.
The target of the MiTM intruders in \textit{UC2}, \textit{UC3}, and \textit{UC4} is the same but they adopt different strategies to accomplish it. %\textcolor{red}{[mention use case 1]}  
These use cases detailed in Section~\ref{use_cases} demonstrate increasing complexity. The time to cause the same overload in branch [5286,  5046] differs based on the strategy in each use case, as illustrated in Fig. ~\ref{fig:flow_impacts}.  For Use Cases~2,~3, and~4, the overload occurs at 173 s, 216 s and 541 s, respectively. The differences in time as well as the system dynamics are due to the amount and sequence of intrusions in these three strategies. 

\subsection{Evaluation of MiTM attack practicality}

%%%%%%%%%%%Let me explain this through action sequence diagram%%%%%%%%%%%%%%%%%%%%%%%%%%%%%%%%%%%%%%

\begin{figure}[!t]
%[h!]
\centerline{\includegraphics[height=2.0 in,width=3.4 in]{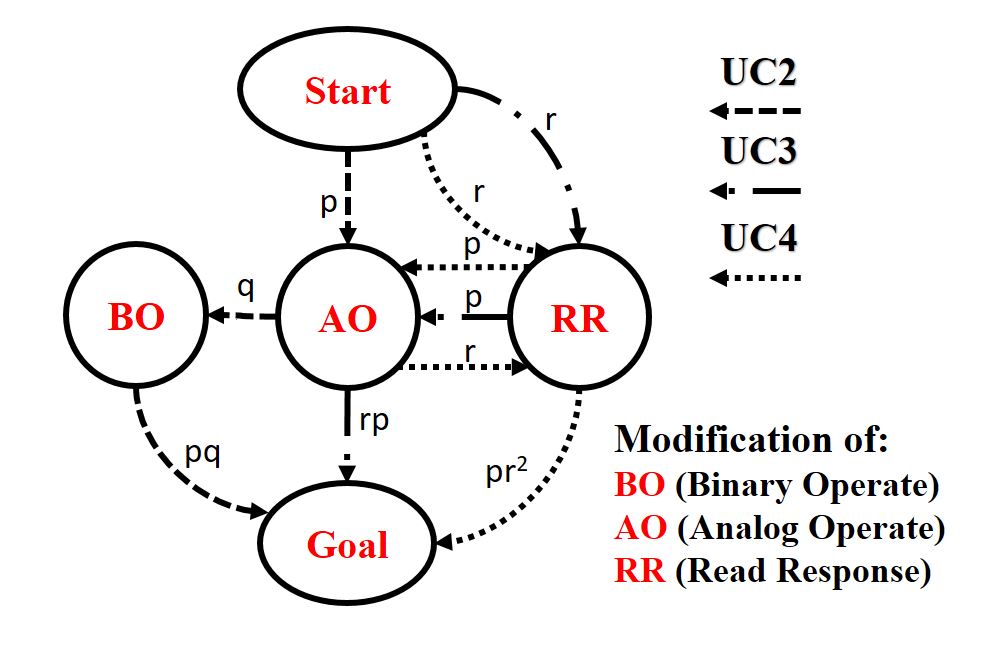}}
\caption{Action sequence of intrusions for \textit{UC2}, \textit{UC3}, and \textit{UC4}.}
\label{fig:intrusion_sequence}
\end{figure}

The successful implementation of the attack use cases requires the intruder to cause the binary operate (BO) and analog operate (AO) FCIs and read response (RR) FDIs in a particular sequence as shown in Fig~\ref{fig:intrusion_sequence}. 
Due to the resource limitations at the attacker, such as sniffing from a single network buffer, it can only accomplish the modification operations with a success probability of $p$, $q$, and $r$ for BO, AO, and RR packets separately. Assuming the number of BO, AO and RR operations to be $m$, $n$ and $o$, the expected number of steps that the intruder has to take to reach its goal is inverse of the success probabilities, which is $p^{m}q^{n}$ for \textit{UC2}, $r^{o}p^{m}$ for \textit{UC3}, and $p^{m}r^{2o}$ for \textit{UC4}.  

The intruder continues the attack until it reaches its goal to overload the branch [5285,5046]. Hence, we evaluate the average minimum number of FCI and FDI modifications the intruder has to perform to reach its target. The success probability depends on the available resources of the intruder, the master polling rate, as well as the number of masters polling the measurements. 

For \textit{UC4}, the number of FDI attempts is higher because the processing time of an FDI is higher than for an FCI, as it involves parsing the DNP3 response from outstations that usually have large payloads.
This higher processing time reduces the success probability of the FDI attack.
For \textit{UC3}, in the RTAC case, an exception of higher FCIs are observed due to the automated generation protection control logics incorporated in the RTAC. The protection logic ensured that the generation setpoint increases when the generation output reading goes below a certain threshold.

Table~\ref{tab:fci_fdi_attempts} shows the minimum number of FCI and FDI attempts on average that are required to accomplish the final goal of the intruder for each use case with both RTAC and OpenDNP3 master.

\begin{table}
    \centering
    \begin{tabular}{|l|l|l|l|l|l|l|}
     \hline
 Type & \multicolumn{3}{|c|}{OpenDNP3 Master} & \multicolumn{3}{|c|}{RTAC Master}\\
 %\hline
    \hline
         & UC2 & UC3 & UC4 & UC2 & UC3 & UC4 \\ \hline
        FDI & N/A & 25.5 & 27.25 & N/A & 17.7 & 30.3 \\ \hline
        FCI & 16.75 & 15.5 & 18.6 & 27.3 & 54.7 & 17.4 \\ \hline
    \end{tabular}
    \caption{Minimum number of FCI and FDI attempts required on\\ average by the intruder for accomplishing its goal in \textit{UC2}, \textit{UC3}, and \textit{UC4}.}
    \label{tab:fci_fdi_attempts}
    \vspace{-0.5 cm}
\end{table}

%%%%%%%%%%%%%The idea is coming from the concept of queueing theory%%%%%%%%%%%%%%%%%%%%%%
From queueing theory, we know that the traffic intensity $\rho$ is computed based on the packet arrival rate $\lambda$ and the service rate $\mu$ as $\rho=\frac{\lambda}{\mu}$~\cite{gallager}. From intruder reference, the arrival rate $\lambda$ is determined by the polling rates from the master as well as the number of DNP3 masters. The service rate $\mu$ is fixed since it is the single intrusion node that processes the incoming traffic. The higher the $\rho$, the lower is the success probability for the intruder to modify the traffic. In literature, the arrival rate distribution can be deterministic or random. In our simulation, since we observe polled traffic as well as commands, it can be considered a random distribution. Every payload that the intruder fails to forward results in drop of the packet, and that triggers retransmissions from the sender.
The algorithms developed and utilized for FDI and FCI attacks affect the processing time.

\subsection{ELK stack visualization}
\begin{figure}[h!]
\centerline{\includegraphics[height=1.9 in,width=3.5 in]{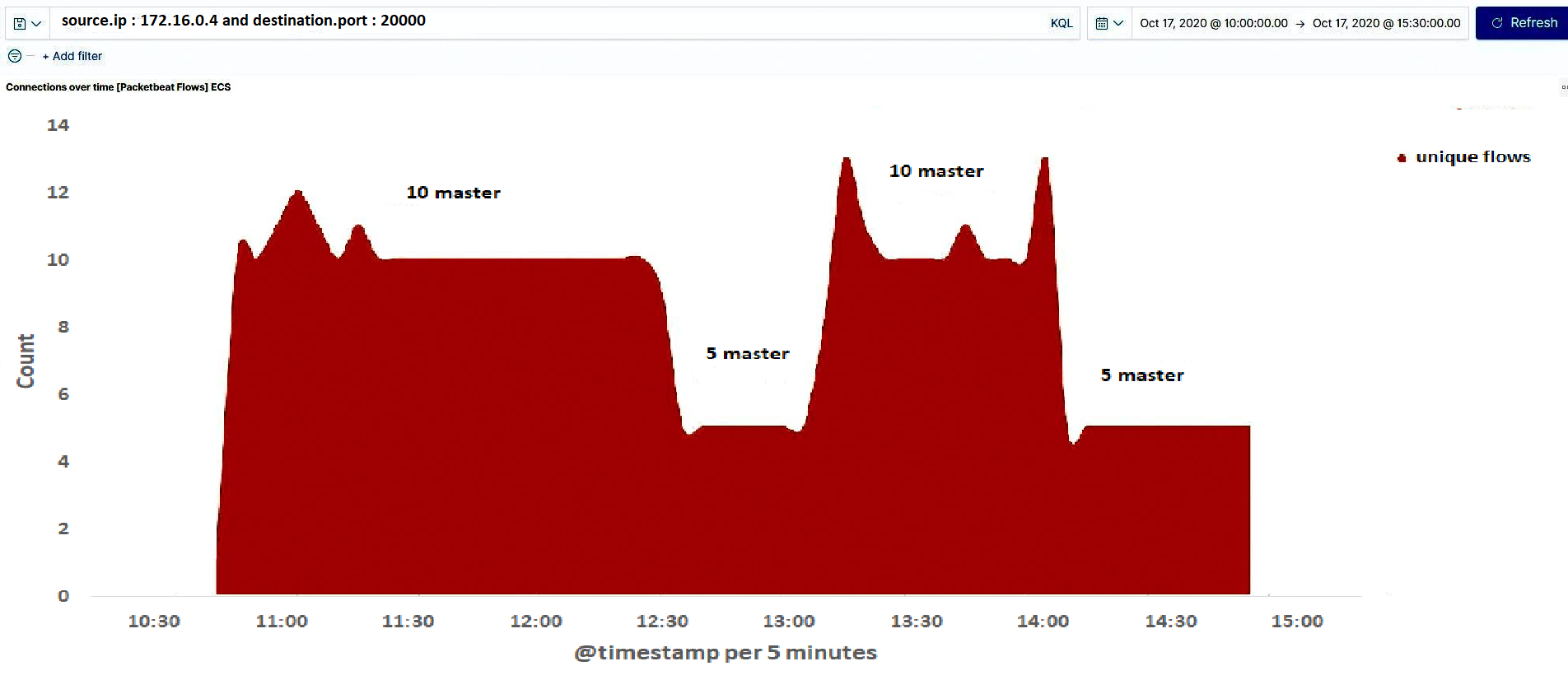}}
\caption{Count of TCP flows from Packetbeat using Kibana while the use cases with 10 and 5 masters are incorporated.}
\label{fig:packetbeat_plot}
\end{figure}

RESLab visualizes the results using ELK stack, where Fig.~\ref{fig:packetbeat_plot} shows a real-time count of the number of active TCP flows while the experiments are being performed for the four use cases with 5 and 10 DNP3 masters. Since the number of active TCP flows is an indicator of the number of connected clients, it helps us to detect if there are more than intended number of clients. At certain times, we observe more than 10 active connections, as some clients lose connection and re-initate a new connection with a different source port number due to the MiTM attacks. 
The number of active TCP flows are the indicator of number of masters connected. Higher variance of connections are observed in 10 master cases due to higher retransmissions.

\begin{figure}[h!]
\centerline{\includegraphics[height=1.9 in,width=3.5 in]{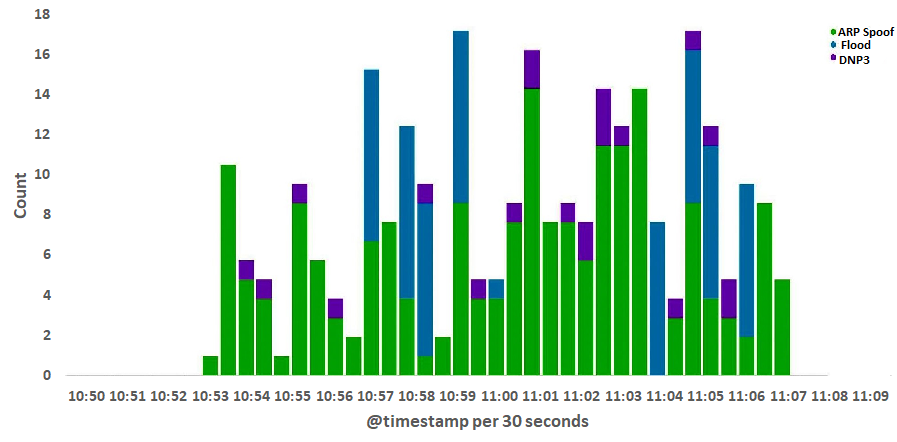}}
\caption{Number of Snort alerts by alert type using Logstash and Kibana.}
\label{fig:logstash_snort}
\end{figure}

The Kibana Query Language (KQL) filters helps us to filter traffic,
%In Fig.~\ref{fig:packetbeat_plot}, we filter 
based on the source IP of the DNP3 master (i.e., 172.16.0.2) and the destination port in the DNP3 outstations (i.e., 20000)
 as shown in Fig.~\ref{fig:packetbeat_plot}. 
 A separate Logstash index is created in Elasticsearch to store real-time snort alerts. Fig.~\ref{fig:logstash_snort} shows the 
 histogram 
 created with Kibana for different types of snort alerts (ICMP flood, ARP spoof, DNP3 operate) during one of the scenarios from the use cases. 
 
\subsection{Discussion}
These results validate the integration of emulators, simulators, hardware, and software tools including visualization and IDS in RESLab by performing 
DoS and MiTM attacks on the power system. 
Through four use cases, RESLab shows how such attacks can impact the electric grid. 

To understand the dynamics of the DoS attack, results 
present the 
impact on RTT and throughput due to different attack intervals and payload sizes of ICMP injections in DoS.
For understanding the dynamics of the MiTM attack, we analyze strategies adopted by the intruder to cause the desired contingencies. We also explore the MiTM attacks with different polling intervals and number of master applications, which impacts on retransmission rates, RTT, and packet processing time. 
The intrusions performed in \textit{UC3} and \textit{UC4} provide the platform to create and mitigate FDI attacks on state estimation which involves an intruder tampering the measurements. 

The simulations performed for the substation network in CORE consisted of one broadcast domain. 
This causes the intruder to observe the traffic related to all the substations.
The number of DNP3 masters are limited to 5 or 10 in our scenarios which is enough to enable the intruder to accomplish its $N-x$ contingencies such that its $x$ components 
are in these 5 or 10 substations. However, they are modeled through a single substation network in CORE. 
Hence, the intruder's capacity to inject modified traffic is resource-limited due to having a single substation LAN in CORE, as the intruder can only process traffic on the single network buffer.

\section{Conclusion}
\label{conclusion}

A cyber-physical testbed provides a platform to understand security threat events and their impact on the power grid. This will help to facilitate grid resiliency to cyber intrusions. In this work, we present our testbed RESLab, where its architecture makes use of components such as vSphere, CORE, PWDS, Snort, among others, to successfully emulate the physical and cyber component of a synthetic large scale electric grid, and we demonstrate the use of DNP3 based control and measurement traffic to and from substation field devices. The methodology and mechanics behind our testbed are demonstrated through four use cases utilizing two types of cyber intrusion experiments: DoS and MiTM. The dynamics of the intrusions are validated by implementing use cases targeting specific parts of the grid. These intrusion events are evaluated from their respective characteristic features, including latency (RTT), throughput, and goodput in the emulated WAN network. 

By providing a safe proving ground for cyber-attack experimentation, RESLab is a platform to study defense mechanisms,
where its ability to generate real-time %traffic and 
datasets 
and customize monitoring, visualization, and detection 
will play a major role
in developing 
cyber-physical state estimation, situational awareness, optimal response, etc.
to prevent impending contingencies.

\section{Acknowledgements}
This research is supported by the US Department of Energy's (DoE) Cybersecurity for Energy Delivery Systems program under award DE-OE0000895.

%%%%%%%%%%%%%%   Bibliography   %%%%%%%%%%%%%%
\normalsize
\bibliography{references}

\begin{thebibliography}{61}
\providecommand{\natexlab}[1]{#1}
\providecommand{\url}[1]{\texttt{#1}}
\expandafter\ifx\csname urlstyle\endcsname\relax
  \providecommand{\doi}[1]{doi: #1}\else
  \providecommand{\doi}{doi: \begingroup \urlstyle{rm}\Url}\fi

\bibitem[{E. Targett}(2020)]{entso}
{E. Targett}.
\newblock {High Voltage Attack: EU’s Power Grid Organisation Hit by Hackers},
  March 2020.
\newblock URL
  \url{https://www.cbronline.com/news/eu-power-grid-organisation-hacked}.

\bibitem[Alert(2016)]{ukraine_2015}
D~Alert.
\newblock Analysis of the cyber attack on the ukrainian power grid, 2016.

\bibitem[Streltsov(2017)]{pivnichna}
Lev Streltsov.
\newblock The system of cybersecurity in ukraine: principles, actors,
  challenges, accomplishments.
\newblock \emph{European Journal for Security Research}, 2\penalty0
  (2):\penalty0 147--184, 2017.

\bibitem[Langner(2011)]{stuxnet}
Ralph Langner.
\newblock Stuxnet: Dissecting a cyberwarfare weapon.
\newblock \emph{IEEE Security \& Privacy}, 9\penalty0 (3):\penalty0 49--51,
  2011.

\bibitem[Pan et~al.(2015)Pan, Morris, and Adhikari]{pan2015developing}
Shengyi Pan, Thomas Morris, and Uttam Adhikari.
\newblock Developing a hybrid intrusion detection system using data mining for
  power systems.
\newblock \emph{IEEE Transactions on Smart Grid}, 6\penalty0 (6):\penalty0
  3104--3113, 2015.

\bibitem[Adhikari et~al.(2016)Adhikari, Morris, and Pan]{adhikari2016wams}
Uttam Adhikari, Thomas Morris, and Shengyi Pan.
\newblock Wams cyber-physical test bed for power system, cybersecurity study,
  and data mining.
\newblock \emph{IEEE Transactions on Smart Grid}, 8\penalty0 (6):\penalty0
  2744--2753, 2016.

\bibitem[Poudel et~al.(2017)Poudel, Ni, and Malla]{poudel2017real}
Shiva Poudel, Zhen Ni, and Naresh Malla.
\newblock Real-time cyber physical system testbed for power system security and
  control.
\newblock \emph{International Journal of Electrical Power \& Energy Systems},
  90:\penalty0 124--133, 2017.

\bibitem[{Yang} et~al.(2017){Yang}, {Xu}, {Gao}, {Yuan}, {McLaughlin}, and
  {Sezer}]{yang_2017}
Y.~{Yang}, H.~{Xu}, L.~{Gao}, Y.~{Yuan}, K.~{McLaughlin}, and S.~{Sezer}.
\newblock Multidimensional intrusion detection system for iec 61850-based scada
  networks.
\newblock \emph{IEEE Transactions on Power Delivery}, 32\penalty0 (2):\penalty0
  1068--1078, 2017.

\bibitem[{Fovino} et~al.(2010){Fovino}, {Masera}, {Guidi}, and
  {Carpi}]{igor_2010}
I.~N. {Fovino}, M.~{Masera}, L.~{Guidi}, and G.~{Carpi}.
\newblock An experimental platform for assessing scada vulnerabilities and
  countermeasures in power plants.
\newblock In \emph{3rd International Conference on Human System Interaction},
  pages 679--686, 2010.

\bibitem[{Oyewumi} et~al.(2019){Oyewumi}, {Jillepalli}, {Richardson},
  {Ashrafuzzaman}, {Johnson}, {Chakhchoukh}, {Haney}, {Sheldon}, and {de
  Leon}]{idaho_testbed}
I.~A. {Oyewumi}, A.~A. {Jillepalli}, P.~{Richardson}, M.~{Ashrafuzzaman}, B.~K.
  {Johnson}, Y.~{Chakhchoukh}, M.~A. {Haney}, F.~T. {Sheldon}, and D.~C. {de
  Leon}.
\newblock Isaac: The idaho cps smart grid cybersecurity testbed.
\newblock In \emph{2019 IEEE Texas Power and Energy Conference (TPEC)}, pages
  1--6, 2019.

\bibitem[{Chen} et~al.(2014){Chen}, {Butler-Purry}, {Goulart}, and
  {Kundur}]{bo_karen}
B.~{Chen}, K.~L. {Butler-Purry}, A.~{Goulart}, and D.~{Kundur}.
\newblock Implementing a real-time cyber-physical system test bed in rtds and
  opnet.
\newblock In \emph{2014 North American Power Symposium (NAPS)}, pages 1--6,
  2014.

\bibitem[{Nelson} et~al.(2016){Nelson}, {Chakraborty}, {Dexin Wang}, {Singh},
  {Qiang Cui}, {Liuqing Yang}, and {Suryanarayanan}]{austin_2016}
A.~{Nelson}, S.~{Chakraborty}, {Dexin Wang}, P.~{Singh}, {Qiang Cui}, {Liuqing
  Yang}, and S.~{Suryanarayanan}.
\newblock Cyber-physical test platform for microgrids: Combining hardware,
  hardware-in-the-loop, and network-simulator-in-the-loop.
\newblock In \emph{2016 IEEE Power and Energy Society General Meeting (PESGM)},
  pages 1--5, 2016.

\bibitem[{Mallouhi} et~al.(2011){Mallouhi}, {Al-Nashif}, {Cox}, {Chadaga}, and
  {Hariri}]{malaz_2011}
M.~{Mallouhi}, Y.~{Al-Nashif}, D.~{Cox}, T.~{Chadaga}, and S.~{Hariri}.
\newblock A testbed for analyzing security of scada control systems (tasscs).
\newblock In \emph{ISGT 2011}, pages 1--7, 2011.

\bibitem[{Aghamolki} et~al.(2015){Aghamolki}, {Miao}, and {Fan}]{hossein_2015}
H.~G. {Aghamolki}, Z.~{Miao}, and L.~{Fan}.
\newblock A hardware-in-the-loop scada testbed.
\newblock In \emph{2015 North American Power Symposium (NAPS)}, pages 1--6,
  2015.

\bibitem[{Sahu} et~al.(2016){Sahu}, {Goulart}, and
  {Butler-Purry}]{abhijeet_testbed}
A.~{Sahu}, A.~{Goulart}, and K.~{Butler-Purry}.
\newblock Modeling ami network for real-time simulation in ns-3.
\newblock In \emph{2016 Principles, Systems and Applications of IP
  Telecommunications (IPTComm)}, pages 1--8, 2016.

\bibitem[{Yafen} et~al.(2011){Yafen}, {Jiaxi}, {Jing}, and {Ning}]{sitl2011}
S.~{Yafen}, C.~{Jiaxi}, Y.~{Jing}, and H.~{Ning}.
\newblock Reliability analysis of system-in-the-loop network platform based on
  delays.
\newblock In \emph{2011 Seventh International Conference on Computational
  Intelligence and Security}, pages 750--753, 2011.

\bibitem[{van Heerden} et~al.(2013){van Heerden}, {Pieterse}, {Burke}, and
  {Irwin}]{vsphere}
R.~{van Heerden}, H.~{Pieterse}, I.~{Burke}, and B.~{Irwin}.
\newblock Developing a virtualised testbed environment in preparation for
  testing of network based attacks.
\newblock In \emph{2013 International Conference on Adaptive Science and
  Technology}, pages 1--8, 2013.

\bibitem[{Tan} et~al.(2012){Tan}, {Song}, {Qifen Dong}, and {Tong}]{core2}
S.~{Tan}, W.~{Song}, {Qifen Dong}, and L.~{Tong}.
\newblock Score: Smart-grid common open research emulator.
\newblock In \emph{2012 IEEE Third International Conference on Smart Grid
  Communications (SmartGridComm)}, pages 282--287, Nov 2012.
\newblock \doi{10.1109/SmartGridComm.2012.6485997}.

\bibitem[{Venkataramanan} et~al.(2016){Venkataramanan}, {Srivastava}, and
  {Hahn}]{core1}
V.~{Venkataramanan}, A.~{Srivastava}, and A.~{Hahn}.
\newblock Real-time co-simulation testbed for microgrid cyber-physical
  analysis.
\newblock In \emph{2016 Workshop on Modeling and Simulation of Cyber-Physical
  Energy Systems (MSCPES)}, pages 1--6, April 2016.
\newblock \doi{10.1109/MSCPES.2016.7480220}.

\bibitem[Raybourn et~al.(2018)Raybourn, Kunz, Fritz, and
  Urias]{raybourn2018zero}
Elaine~M Raybourn, Michael Kunz, David Fritz, and Vince Urias.
\newblock A zero-entry cyber range environment for future learning ecosystems.
\newblock In \emph{Cyber-Physical Systems Security}, pages 93--109. Springer,
  2018.

\bibitem[Johnson(2017)]{sceptre}
Jay Johnson.
\newblock Sceptre: Power system and networking co-simulation environment, 07
  2017.

\bibitem[{Palmintier} et~al.(2017){Palmintier}, {Krishnamurthy}, {Top},
  {Smith}, {Daily}, and {Fuller}]{helics}
B.~{Palmintier}, D.~{Krishnamurthy}, P.~{Top}, S.~{Smith}, J.~{Daily}, and
  J.~{Fuller}.
\newblock Design of the helics high-performance
  transmission-distribution-communication-market co-simulation framework.
\newblock In \emph{2017 Workshop on Modeling and Simulation of Cyber-Physical
  Energy Systems (MSCPES)}, pages 1--6, 2017.

\bibitem[Hong et~al.(2015)Hong, Nuqui, Ishchenko, Wang, Cui, Kondabathini,
  Coats, and Kunsman]{hong2015cyber}
Junho Hong, Reynaldo Nuqui, Dmitry Ishchenko, Zhenyuan Wang, Tao Cui, Anil
  Kondabathini, David Coats, and S~Kunsman.
\newblock Cyber-physical security test bed: A platform for enabling
  collaborative cyber defense methods.
\newblock In \emph{PACWorld Americas Conference}, 2015.

\bibitem[Liu et~al.(2015)Liu, Vellaithurai, Biswas, Gamage, and
  Srivastava]{liu2015analyzing}
Ren Liu, Ceeman Vellaithurai, Saugata~S Biswas, Thoshitha~T Gamage, and
  Anurag~K Srivastava.
\newblock Analyzing the cyber-physical impact of cyber events on the power
  grid.
\newblock \emph{IEEE Transactions on Smart Grid}, 6\penalty0 (5):\penalty0
  2444--2453, 2015.

\bibitem[Kezunovic et~al.(2017)Kezunovic, Esmailian, Govindarasu, and
  Mehrizi-Sani]{kezunovic2017use}
Mladen Kezunovic, Ahad Esmailian, Manimaran Govindarasu, and Ali Mehrizi-Sani.
\newblock The use of system in the loop, hardware in the loop, and co-modeling
  of cyber-physical systems in developing and evaluating new smart grid
  solutions.
\newblock In \emph{Proceedings of the 50th Hawaii International Conference on
  System Sciences}, 2017.

\bibitem[Ashok et~al.(2015)Ashok, Wang, Brown, and
  Govindarasu]{ashok2015experimental}
Aditya Ashok, Pengyuan Wang, Matthew Brown, and Manimaran Govindarasu.
\newblock Experimental evaluation of cyber attacks on automatic generation
  control using a cps security testbed.
\newblock In \emph{2015 IEEE Power \& Energy Society General Meeting}, pages
  1--5. IEEE, 2015.

\bibitem[Papaspiliotopoulos et~al.(2015)Papaspiliotopoulos, Korres, Kleftakis,
  and Hatziargyriou]{papaspiliotopoulos2015hardware}
Vasileios~A Papaspiliotopoulos, George~N Korres, Vasilis~A Kleftakis, and
  Nikos~D Hatziargyriou.
\newblock Hardware-in-the-loop design and optimal setting of adaptive
  protection schemes for distribution systems with distributed generation.
\newblock \emph{IEEE Transactions on Power Delivery}, 32\penalty0 (1):\penalty0
  393--400, 2015.

\bibitem[{Ashok} et~al.(2015){Ashok}, {Pengyuan Wang}, {Brown}, and
  {Govindarasu}]{iowaTestbed}
A.~{Ashok}, {Pengyuan Wang}, M.~{Brown}, and M.~{Govindarasu}.
\newblock Experimental evaluation of cyber attacks on automatic generation
  control using a cps security testbed.
\newblock In \emph{2015 IEEE Power Energy Society General Meeting}, pages 1--5,
  2015.

\bibitem[{Overbye} et~al.(2019){Overbye}, {Mao}, {Birchfield}, {Weber}, and
  {Davis}]{DS}
T.~J. {Overbye}, Z.~{Mao}, A.~{Birchfield}, J.~D. {Weber}, and M.~{Davis}.
\newblock {An Interactive, Stand-Alone and Multi-User Power System Simulator
  for the PMU Time Frame}.
\newblock In \emph{2019 IEEE Texas Power and Energy Conference (TPEC)}, pages
  1--6, 2019.

\bibitem[Overbye et~al.(2017)Overbye, Mao, Shetye, and
  Weber]{overbye2017interactive}
Thomas~J Overbye, Zeyu Mao, Komal~S Shetye, and James~D Weber.
\newblock An interactive, extensible environment for power system simulation on
  the pmu time frame with a cyber security application.
\newblock In \emph{2017 IEEE Texas Power and Energy Conference (TPEC)}, pages
  1--6. IEEE, 2017.

\bibitem[Mao et~al.(2020)Mao, Huang, and Davis]{w4ips}
Zeyu Mao, Hao Huang, and Katherine Davis.
\newblock W4ips: A web-based interactive power system simulation environment
  for power system security analysis.
\newblock In \emph{Proceedings of the 53rd Hawaii International Conference on
  System Sciences}, 2020.

\bibitem[Johnson et~al.(2018)Johnson, Ablinger, Bruendlinger, Fox, and
  Flicker]{johnson2018interconnection}
Jay Johnson, Ron Ablinger, Roland Bruendlinger, Bob Fox, and Jack Flicker.
\newblock Interconnection standard grid-support function evaluations using an
  automated hardware-in-the-loop testbed.
\newblock \emph{IEEE Journal of Photovoltaics}, 8\penalty0 (2):\penalty0
  565--571, 2018.

\bibitem[Thornton et~al.(2017)Thornton, Motalleb, Smidt, Branigan, Siano, and
  Ghorbani]{thornton2017internet}
Matsu Thornton, Mahdi Motalleb, Holm Smidt, John Branigan, Pierluigi Siano, and
  Reza Ghorbani.
\newblock Internet-of-things hardware-in-the-loop simulation architecture for
  providing frequency regulation with demand response.
\newblock \emph{IEEE Transactions on Industrial Informatics}, 14\penalty0
  (11):\penalty0 5020--5028, 2017.

\bibitem[Piesciorovsky and Schulz(2017)]{piesciorovsky2017fuse}
Emilio~C Piesciorovsky and Noel~N Schulz.
\newblock Fuse relay adaptive overcurrent protection scheme for microgrid with
  distributed generators.
\newblock \emph{IET Generation, Transmission \& Distribution}, 11\penalty0
  (2):\penalty0 540--549, 2017.

\bibitem[{Becejac} et~al.(2020){Becejac}, {Eppinger}, {Ashok}, {Agrawal}, and
  {O'Brien}]{tamara_pnnl}
T.~{Becejac}, C.~{Eppinger}, A.~{Ashok}, U.~{Agrawal}, and J.~{O'Brien}.
\newblock Prime: a real-time cyber-physical systems testbed: from wide-area
  monitoring, protection, and control prototyping to operator training and
  beyond.
\newblock \emph{IET Cyber-Physical Systems: Theory Applications}, 5\penalty0
  (2):\penalty0 186--195, 2020.

\bibitem[{Azimian} et~al.(2019){Azimian}, {Adhikari}, {Vanfretti}, and
  {Hooshyar}]{behrouz_2019}
B.~{Azimian}, P.~M. {Adhikari}, L.~{Vanfretti}, and H.~{Hooshyar}.
\newblock Cross-platform comparison of standard power system components used in
  real time simulation.
\newblock In \emph{2019 7th Workshop on Modeling and Simulation of
  Cyber-Physical Energy Systems (MSCPES)}, pages 1--6, 2019.

\bibitem[Stifter et~al.(2018)Stifter, Cordova, Kazmi, and
  Arghandeh]{stifter2018real}
Matthias Stifter, Jose Cordova, Jawad Kazmi, and Reza Arghandeh.
\newblock Real-time simulation and hardware-in-the-loop testbed for
  distribution synchrophasor applications.
\newblock \emph{Energies}, 11\penalty0 (4):\penalty0 876, 2018.

\bibitem[{Yang} et~al.(2012){Yang}, {McLaughlin}, {Littler}, {Sezer}, {Im},
  {Yao}, {Pranggono}, and {Wang}]{yang_2012}
Y.~{Yang}, K.~{McLaughlin}, T.~{Littler}, S.~{Sezer}, E.~G. {Im}, Z.~Q. {Yao},
  B.~{Pranggono}, and H.~F. {Wang}.
\newblock Man-in-the-middle attack test-bed investigating cyber-security
  vulnerabilities in smart grid scada systems.
\newblock In \emph{International Conference on Sustainable Power Generation and
  Supply (SUPERGEN 2012)}, pages 1--8, 2012.

\bibitem[Kundu et~al.(2020)Kundu, Sahu, Serpedin, and Davis]{se_attack}
Arnav Kundu, Abhijeet Sahu, Erchin Serpedin, and Katherine Davis.
\newblock A3d: Attention-based auto-encoder anomaly detector for false data
  injection attacks.
\newblock \emph{Electric Power Systems Research}, 189:\penalty0 106795, 2020.
\newblock ISSN 0378-7796.
\newblock \doi{https://doi.org/10.1016/j.epsr.2020.106795}.
\newblock URL
  \url{http://www.sciencedirect.com/science/article/pii/S0378779620305988}.

\bibitem[Ozay et~al.(2016)Ozay, Esnaola, Vural, Kulkarni, and
  Poor]{ozay2016machine}
Mete Ozay, Inaki Esnaola, Fatos Tunay~Yarman Vural, Sanjeev~R Kulkarni, and
  H~Vincent Poor.
\newblock Machine learning methods for attack detection in the smart grid.
\newblock \emph{IEEE Transactions on Neural Networks and Learning Systems},
  27\penalty0 (8):\penalty0 1773--1786, 2016.

\bibitem[ope()]{open_dnp3}
Open dnp3 documentation.
\newblock \url{https://dnp3.github.io/}.

\bibitem[cyb(2016)]{cyber_killchain}
Analysis of the cyber attack on the ukrainian power grid: Defense use case.
\newblock \url{https://ics.sans.org/media/E-ISAC_SANS_Ukraine_DUC_5.pdf}, 2016.

\bibitem[Birchfield et~al.(2017)Birchfield, Xu, Gegner, Shetye, and
  Overbye]{synthetic}
A.~B. Birchfield, T.~Xu, K.~M. Gegner, K.~S. Shetye, and T.~J. Overbye.
\newblock Grid structural characteristics as validation criteria for synthetic
  networks.
\newblock \emph{IEEE Transactions on Power Systems}, 32\penalty0 (4), July
  2017.
\newblock ISSN 0885-8950.

\bibitem[{Wlazlo} et~al.(2019){Wlazlo}, {Price}, {Veloz}, {Sahu}, {Huang},
  {Goulart}, {Davis}, and {Zounouz}]{synthetic_comm}
P.~{Wlazlo}, K.~{Price}, C.~{Veloz}, A.~{Sahu}, H.~{Huang}, A.~{Goulart},
  K.~{Davis}, and S.~{Zounouz}.
\newblock A cyber topology model for the texas 2000 synthetic electric power
  grid.
\newblock In \emph{2019 Principles, Systems and Applications of IP
  Telecommunications (IPTComm)}, pages 1--8, 2019.

\bibitem[{Davis} et~al.(2006){Davis}, {Tate}, {Okhravi}, {Grier}, {Overbye},
  and {Nicol}]{nicol_overbye}
C.~M. {Davis}, J.~E. {Tate}, H.~{Okhravi}, C.~{Grier}, T.~J. {Overbye}, and
  D.~{Nicol}.
\newblock Scada cyber security testbed development.
\newblock In \emph{2006 38th North American Power Symposium}, pages 483--488,
  2006.

\bibitem[{Gaudet} et~al.(2020){Gaudet}, {Sahu}, {Goulart}, {Rogers}, and
  {Davis}]{firewall_paper}
N.~{Gaudet}, A.~{Sahu}, A.~E. {Goulart}, E.~{Rogers}, and K.~{Davis}.
\newblock Firewall configuration and path analysis for smartgrid networks.
\newblock In \emph{2020 IEEE International Workshop Technical Committee on
  Communications Quality and Reliability (CQR)}, pages 1--6, 2020.

\bibitem[Huang and Davis(2018)]{huang2018extracting}
Hao Huang and Katherine Davis.
\newblock Extracting substation cyber-physical architecture through intelligent
  electronic devices' data.
\newblock In \emph{2018 IEEE Texas Power and Energy Conference (TPEC)}, pages
  1--6. IEEE, 2018.

\bibitem[Schroder(2018)]{quagga}
Carla Schroder.
\newblock Dynamic linux routing with quagga.
\newblock
  \url{https://www.linux.com/topic/networking/dynamic-linux-routing-quagga/},
  2018.

\bibitem[cor()]{core_ospf}
Core services.
\newblock \url{http://coreemu.github.io/core/services.html}.

\bibitem[rta(2018)]{rtac_acselerator}
Acselerator rtac sel-5033 software instruction manual.
\newblock \url{https://selinc.com/products/5033/}, 2018.

\bibitem[East et~al.(2009)East, Butts, Papa, and Shenoi]{dnp3_na}
Samuel East, Jonathan Butts, Mauricio Papa, and Sujeet Shenoi.
\newblock A taxonomy of attacks on the dnp3 protocol.
\newblock volume 311, 03 2009.
\newblock \doi{10.1007/978-3-642-04798-5_5}.

\bibitem[{S}chweitzer {E}ngineering~{L}aboratories. {I}nc(2020)]{sel_2020}
{S}chweitzer {E}ngineering~{L}aboratories. {I}nc, Oct 2020.
\newblock URL \url{https://selinc.com/products/3530/}.

\bibitem[Orebaugh et~al.(2005)Orebaugh, Biles, and Babbin]{snort_cookbook}
Angela~D. Orebaugh, Simon Biles, and Jacob Babbin.
\newblock \emph{Snort Cookbook}.
\newblock O’Reilly Media, Inc., 2005.
\newblock ISBN 0596007914.

\bibitem[pac()]{packetbeat}
Packetbeat in elk stack.
\newblock \url{https://www.elastic.co/beats/packetbeat}.

\bibitem[zab()]{zabbix}
Zabbix for network monitoring.
\newblock \url{https://www.zabbix.com/network_monitoring}.

\bibitem[{Ortega} et~al.(2009){Ortega}, {Marcos}, {Chiang}, and
  {Abad}]{arp_cache_poison}
A.~P. {Ortega}, X.~E. {Marcos}, L.~D. {Chiang}, and C.~L. {Abad}.
\newblock Preventing arp cache poisoning attacks: A proof of concept using
  openwrt.
\newblock In \emph{2009 Latin American Network Operations and Management
  Symposium}, pages 1--9, 2009.
\newblock \doi{10.1109/LANOMS.2009.5338799}.

\bibitem[Narimani et~al.(2020)Narimani, Huang, Umunnakwe, Mao, Sahu, Zonouz,
  and Davis]{n_x}
Mohammad Narimani, Hao Huang, Amarachi Umunnakwe, Zeyu Mao, Abhijeet Sahu,
  Saman Zonouz, and Kate Davis.
\newblock Generalized contingency analysis based on graph theory and line
  outage distribution factor, 07 2020.

\bibitem[{Kalluri} et~al.(2016){Kalluri}, {Mahendra}, {Kumar}, and
  {Prasad}]{dos_scada}
R.~{Kalluri}, L.~{Mahendra}, R.~K.~S. {Kumar}, and G.~L.~G. {Prasad}.
\newblock Simulation and impact analysis of denial-of-service attacks on power
  scada.
\newblock In \emph{2016 National Power Systems Conference (NPSC)}, pages 1--5,
  2016.
\newblock \doi{10.1109/NPSC.2016.7858908}.

\bibitem[Day(2015)]{dnp3_crob}
Timothy Day.
\newblock Dnp3, distributed network protocol v3 an introduction.
\newblock
  \url{https://na.eventscloud.com/file_uploads/b68188f3ce5b22895a67b1afe5e51b6a_DNP3IntroductionHORS.PDF},
  2015.

\bibitem[Sanfilippo()]{hping3}
Salvatore Sanfilippo.
\newblock Dos attack tool.
\newblock URL \url{http://www.hping.org/hping3.html}.

\bibitem[Bertsekas and Gallager(1992)]{gallager}
Dimitri Bertsekas and Robert Gallager.
\newblock \emph{Data Networks (2nd Ed.)}.
\newblock Prentice-Hall, Inc., USA, 1992.
\newblock ISBN 0132009161.

\end{thebibliography}

%%%%%%%%%%%%  Supplementary Figures  %%%%%%%%%%%%
%\clearpage

%%%%%%%%%%%%%%%%   End   %%%%%%%%%%%%%%%%
%\end{multicols}  % Method B for two-column formatting (doesn't play well with line numbers), comment out if using method A
\end{document}